\documentclass[a4paper,12pt]{article}

\usepackage[usenames]{color}
\usepackage{graphicx}
\usepackage{amsmath}
\usepackage{amssymb}

\newcommand {\cN}{{\cal N}}

\newcommand {\cV}{{\cal V}}

\def\a{\alpha}

\def\z{\zeta}

\def\U{\Upsilon}

\makeatletter
\@addtoreset{equation}{section}
\makeatother

\newcommand{\be}{\begin{equation}}
\newcommand{\ee}{\end{equation}}
\newcommand{\bea}{\begin{eqnarray}}
\newcommand{\eea}{\end{eqnarray}}
\newcommand{\non}{\nonumber}
\def\double #1{#1{\hbox{\kern-2pt $#1$}}}

\begin{document}
\begin{titlepage}
\null
\begin{flushright}
December, 2011
\end{flushright}

\vskip 1.8cm
\begin{center}
 
  {\Large \bf Off-shell Construction of\\ 
\vskip 0.2cm
Superconformal Chern-Simons Theories \\
\vskip 0.3cm
 in Three Dimensions}

\vskip 1.8cm
\normalsize

  {\bf Masato Arai$^{\dagger}$\footnote{masato.arai(at)utef.cvut.cz} 
and Shin Sasaki$^\sharp$\footnote{shin-s(at)kitasato-u.ac.jp}}

\vskip 0.5cm

  { \it 
  $^\dagger$Institute of Experimental and Applied Physics, \\
   Czech Technical University in Prague, \\
   Horsk\' a 3a/22, 128 00, Prague 2, Czech Republic\\
    \vskip 0.5cm
  $^\sharp$Department of Physics \\
  Kitasato University \\
  Sagamihara, 228-8555, Japan
  }

\vskip 2cm

\begin{abstract}
We propose an off-shell construction of 
 three-dimensional $\mathcal{N} = 3$ and $\mathcal{N} = 4$ 
 superconformal Abelian Chern-Simons theories in the projective
 superspace formalism. We also construct coupling 
terms among the gauge fields and matter hypermultiplets.
\end{abstract}

\end{center}
\end{titlepage}

\newpage

\section{Introduction}
Chern-Simons theories in three dimensions have been attracted 
physicist's attention because of its importance in condensed matter and particle physics.
Especially in the field of the latter, supersymmetric extensions of
Chern-Simons theories have been intensively studied. 
It is known that three-dimensional $\mathcal{N} = 2$
Chern-Simons-matter models admit (non)topological solitons 
due to the non-trivial Higgs potential determined by the $\mathcal{N} = 2$ supersymmetry.
The formulation of Chern-Simons-matter models with 
$\mathcal{N} = 3$ supersymmetry is examined in \cite{KaLe}. 
The authors showed that the maximal supersymmetry of Chern-Simons-matter 
models in three dimensions with a single gauge field and no gravity is 
$\mathcal{N} = 3$.
Even though the $\mathcal{N} = 3$ supersymmetry is the maximal one 
in that case, $\mathcal{N} \ge 4$ supersymmetries are possible for 
pure Chern-Simons theories \cite{NiGa,NiGa2,BrGa} and quiver gauge theories.

Recently the low-energy effective theory of multiple 
M2-branes is proposed by Bagger, Lambert and Gustavsson (BLG
model) \cite{BaLa,Gu} which is based on the idea of the novel gauge group
$\mathcal{A}_4$ constructed by 3-algebras. 
Soon after the proposal, it is shown that the BLG model with
$\mathcal{A}_4$ group is nothing but the $\mathcal{N} = 8$ superconformal Chern-Simons-matter model
with gauge group $SU(2) \times SU(2)$ with bi-fundamental matters
\cite{Ra}. Other Chern-Simons models with products of gauge groups and
matters, such as $\mathcal{N} =4$ and $\mathcal{N} = 5$ superconformal 
Chern-Simons-matter models \cite{HoLeLeLePa}, the $\mathcal{N} = 6$ $U(N)
\times U(N)$ model (ABJM model) \cite{AhBeJaMa} have also been constructed. 

Besides these facts, manifestly supersymmetric formulations of
Chern-Simons-matter models have been interesting topics.
For example, Abelian and non-Abelian Chern-Simons-matter models
in three-dimensional $\mathcal{N} = 2$ superspace 
are constructed in \cite{Iv,GaNi1}. 
It is known that to incorporate manifest and off-shell $\mathcal{N} \ge 3$
supersymmetries (hence $\mathcal{N} \ge 2$ in four dimensions), the
ordinary superspace approach is not suitable.  
The on-shell superfield formulation of the $\mathcal{N} = 8$ BLG and
the $\mathcal{N} = 6$ ABJM models are found in \cite{Ce}. 
A good way to introduce the off-shell $\mathcal{N} \ge 3$
supersymmetries is to use the harmonic superspace approach
\cite{GaIvOgSo, GaIvKaOgSo}. 
Pure (without matter) $\mathcal{N} = 5, 6$ Chern-Simons theories 
are studied in the framework of harmonic superspace \cite{HoLe, Zu}.
For the Chern-Simons theories with matter fields, 
a manifestly supersymmetric construction of the $\mathcal{N} = 6$
ABJM model is investigated in the  $\mathcal{N} = 3$ harmonic superspace
\cite{BuIvLePlSaZu}. 

Another way to treat the off-shell extended supersymmetries 
is to use the projective superspace approach \cite{KaLiRo, LR1, LiRo}
which keeps $\mathcal{N} = 2$ manifest supersymmetry in four dimensions. 
The two approaches have the relationship \cite{Ku}
and it is quite interesting to investigate 
the manifestly supersymmetric formulation
of Chern-Simons-matter models in the projective superspace. 
Even more, it is possible to construct the action with manifest
superconformal invariance in the projective superspace \cite{KuPaTaUn, KuLiTa}.
In this paper, we study manifest 
$\mathcal{N} = 3$ and $\mathcal{N} = 4$ superconformal formulations of
Chern-Simons-matter models in the projective superspaces. The analysis of
this paper provides an alternative formulation of supersymmetric
Chern-Simons theories other than the harmonic superspace approach.
A projective superspace formulation of the BLG model is proposed in 
\cite{ChDoSa}. We will give a comment on this construction in the discussion.

The organization of this paper is as follows. 
In the next section, 
the $\mathcal{N} = 3$ supersymmetric Chern-Simons-matter models in
the $\mathcal{N} = 2$ superspace is shown. 
In Section 3, we give a brief review of the three-dimensional
$\mathcal{N} = 3$ projective superspace formulation of superconformal theories.
In Section 4, we formulate the Abelian Chern-Simons-matter models in
$\mathcal{N} = 3$ projective superspace approach and show that the
proposed action correctly reproduces the result constructed in
$\mathcal{N} = 2$ superspace. 
Section 5 is devoted to the $\mathcal{N} = 4$ generalization of the construction. 
Section 6 is conclusion and discussions where 
non-Abelian generalization is briefly discussed.
Notations and conventions of $\mathcal{N} = 2,3,4$ superspaces are given in
Appendix A. 
Detail calculations of the solution to the projective superspace
constraint are shown in Appendix B
The anti-commutation relations among the gauge covariant
derivatives are found in Appendix C.

\section{Chern-Simons-matter models in $\mathcal{N}=2$ superspace}
In this section, we briefly introduce the 
three-dimensional Chern-Simons-matter models in $\mathcal{N} = 2$ superspace.
It has been shown that Chern-Simons-matter models with a single gauge group
have $\mathcal{N} = 3$ maximal supersymmetry \cite{KaLe, Sc} which will be 
enhanced to $\mathcal{N} > 3$ superconformal symmetries when 
appropriate gauge groups and matters are added \cite{HoLeLeLePa, AhBeJaMa}. 

Although our main interest is Abelian Chern-Simons-matter models, 
we start from non-Abelian gauge groups for generality.
We first consider an $\mathcal{N}=2$ $U(N)$ Chern-Simons-matter model 
with level $k$, interacting with $N_f$ flavors.
The model consists of the three-dimensional $\mathcal{N} = 2$ vector
superfield $V_0$ and the 
chiral, antichiral superfields $Q_i ,\ \bar{Q}_i \ (i = 1, \cdots N_f)$ representing the gauge field 
and matters respectively. 
The chiral and antichiral superfields which satisfy the conditions $\bar{\mathbb{D}}_{\alpha} Q_i = 0$
and $\mathbb{D}_{\alpha}\bar{Q}_i=0$ are expanded as 
\begin{eqnarray}
\begin{aligned}
& Q_i (x_L, \theta) = q_i (x_L) + \sqrt{2} \theta \psi_{q i} (x_L) 
+ \theta^2 F_{q i} (x_L), \\
& \bar{Q}_i(x_R,\bar{\theta}) = 
\bar{q}_i(x_R) 
- 
\sqrt{2}\bar{\theta} \bar{\psi}_{\bar{q}i}(x_R)
-
\bar{\theta}^2 \bar{F}_{\bar{q}i}(x_R),
\end{aligned}
\end{eqnarray}
where $\bar{\mathbb{D}}_{\alpha}$ and $\mathbb{D}_\alpha$ are the supercovariant derivatives
in the $\mathcal{N} = 2$ superspace, 
$x_L$ and $x_R$ are the chiral and antichiral coordinates defined in Appendix A.
The vector superfield in the Wess-Zumino gauge is expanded as 
\begin{eqnarray}
V_0 (x, \theta, \bar{\theta}) = 2 i \theta \bar{\theta} \sigma (x) + 2 \theta 
\gamma^{m} \bar{\theta} A_{m} (x) - \sqrt{2} i \bar{\theta}^2 \theta 
\chi (x) + \sqrt{2} i \theta^2 \bar{\theta} \bar{\chi} (x) + \theta^2 \bar{\theta}^2 D (x).
\end{eqnarray}
Here $A_{m} \ (m=0,1,2)$ is the gauge field, $\chi$ is the gaugino, $D$ is the 
auxiliary field and $\sigma$ is the real scalar.
All the component fields in $V_0$ are in the adjoint representation of the gauge 
group. This is obtained by the dimensional reduction of the 
four-dimensional $\mathcal{N} = 1$ vector superfield. 
Actually, $\sigma$ is the $A_3$ 
component of the four-dimensional gauge field.
In order to write down the $\mathcal{N} = 2$ supersymmetric action, 
we employ the trick proposed in \cite{Iv} by 
introducing an auxiliary integration variable $t$ and express the 
Chern-Simons part as the integration of the exponentiated vector superfield. 
The action is given by 
\begin{eqnarray}
S^{\mathcal{N} = 2}_{\mathrm{CSH}} = \int \! d^3 x \int \! d^4 \theta 
\left\{-\frac{ik}{4\pi} 
\int^1_0 \! dt \ \mathrm{Tr} 
\left[
V_0 \bar{\mathbb{D}}^{\alpha} \left( e^{-t V_0} \mathbb{D}_{\alpha} e^{t V_0} \right)
\right]
+ \sum_{i=1}^{N_f} \bar{Q}_i e^{V_0} Q_i
\right\},
\label{N2CSH}
\end{eqnarray}
where the symbol $\mathrm{Tr}$ is the gauge trace
and $Q_i$ ($\bar{Q}_i$)
 are in 
the (anti)fundamental 
representation of the gauge group.
The gauge trace is normalized as $\mathrm{Tr} (T^a T^b) = \delta^{ab}$ 
for the $U(N)$ generators $T^a$.
The Chern-Simons level $k$ should be quantized to be integer
valued for the gauge group $U(N)$. 
The action is gauge invariant under the 
following gauge transformation, 
\begin{eqnarray}
e^{V_0} \longrightarrow e^{ i \bar{\Lambda}} e^{V_0} e^{- i \Lambda}, \qquad 
Q_i \to e^{ i \Lambda} Q_i,\qquad \bar{Q}_i \to e^{-i\bar{\Lambda}} \bar{Q}_i,
\end{eqnarray}
where $\Lambda$, $\bar{\Lambda}$ are gauge parameters satisfying the
chiral, antichiral superfield conditions respectively. 
The 
second term in \eqref{N2CSH} is obtained just by the dimensional reduction 
of the four-dimensional 
matter kinetic term 
while the first term in (\ref{N2CSH}) can 
not be obtained from the four dimensions.
Note that this model is quantum mechanically conformal provided that 
there is no superpotential for $Q_i$ \cite{GaYi}.

Next, we consider $\mathcal{N} = 3$ Chern-Simons-matter theories.
To construct the $\mathcal{N} = 3$ supersymmetric action in
$\mathcal{N} = 2$ superfield formalism, 
one needs to introduce other chiral and antichiral multiplets 
$\Phi, \bar{\Phi}$ with adjoint representation of the gauge group.
They are non-dynamical auxiliary fields. Combined with the vector 
superfield $V_0$, these form the $\mathcal{N} = 4$ vector multiplet. 
The vector multiplet can couple to $\mathcal{N}=4$ hypermultiplets which 
are represented by pairs of chiral and antichiral multiplets $(S_i,
T_i)$, $(\bar{S}_i, \bar{T}_i)$ transforming in 
conjugate representations of the gauge group. 
The system including the Chern-Simons part and the matter part
has only $\mathcal{N}=3$ supersymmetry since the Chern-Simons term
breaks $\mathcal{N}=4$ supersymmetry down to $\mathcal{N}=3$.
The $\mathcal{N} = 3$ supersymmetric Chern-Simons action is given by \cite{GaYi}
\begin{eqnarray}
\begin{aligned}
S^{\mathcal{N} = 3}_{\mathrm{CS}} &= 
\frac{-ik}{4\pi}\int \! d^3 x d^4 \theta \int^1_0 \! d t \ 
\mathrm{Tr}
\left[
V_0 \bar{\mathbb{D}}^{\alpha} \left( e^{t V_0} \mathbb{D}_{\alpha} e^{-t V_0} \right)
\right]
\\
& \qquad 
- \frac{k}{4\pi} \int \! d^3 x d^2 \theta \ \mathrm{Tr} \Phi^2 
+ 
\frac{k}{4\pi} \int \! d^3 x d^2 \bar{\theta} \ \mathrm{Tr}
 \bar{\Phi}^2,
\end{aligned}
\label{N3CS}
\end{eqnarray}
while the matter part is 
\begin{eqnarray}
S^{\mathcal{N} = 3}_{\mathrm{H}} = 
\int \! d^3 x d^4 \theta \sum_{i=1}^{N_f} 
\left(
\bar{S}_i e^{V_0} S_i + T_i e^{-V_0} \bar{T}_i
\right) + 
\sum_{i=1}^{N_f}
\left[ 
2 \! \int \! d^3 x d^2 \theta \ T_i \Phi S_i 
- 2 \! \int \! d^3 x d^2 \bar{\theta} \ \bar{T}_i \bar{\Phi} \bar{S}_i 
\right].
\label{N3higgs}
\end{eqnarray}
The action $S^{\mathcal{N} = 3}_{\mathrm{CS}} + S^{\mathcal{N} = 
3}_{\mathrm{H}}$ keeps only $\mathcal{N} = 2$ {\it manifest} supersymmetry but 
actually preserves $\mathcal{N} = 3$ supersymmetry. This model is also quantum 
mechanically conformal. 

Finally, let us take the Abelian limit of (\ref{N3CS}) for later convenience. 
In the Abelian case, we do not need the auxiliary $t$-integration. Therefore 
the $\mathcal{N} = 3$ Abelian Chern-Simons action is given by 
\begin{eqnarray}
S^{\mathcal{N} = 3}_{\rm CS} = 
\frac{ik}{8\pi}
\int \! d^3 x d^4 \theta \ 
V_0 \bar{\mathbb{D}}^{\alpha} \mathbb{D}_{\alpha} V_0
- \frac{k}{4\pi}
\int \! d^3 x d^2 \theta \ \Phi^2 
+ 
\frac{k}{4\pi}
\int \! d^3 x d^2 \bar{\theta} \ \bar{\Phi}^2.
\label{N2CSHU1}
\end{eqnarray}

\section{Brief survey of projective superspace formalism}
In this section, we briefly review the basic ideas of the three-dimensional $\mathcal{N} = 3$
superconformal projective superspace \cite{KuPaTaUn}.
For those who are not familiar with 
the projective superspace approach, let us
recall the ordinary $d=4, \mathcal{N} = 1$ superfield formalism.
The $d=4$, $\mathcal{N}=1$ superspace is 
parametrized by the space-time coordinate $x^{\mu} \ (\mu = 0, \cdots, 3)$ and $SO(1,3)$
spinor coordinates $\theta_{\alpha}, \bar{\theta}_{\dot{\alpha}}$.
As an explicit example, we consider a supersymmetric Lagrangian constructed by chiral superfields.
A chiral superfield $\Phi$ is not a function of the full superspace, but a function
of its subspace, called the chiral subspace.
This subspace is defined by the constraint 
$\bar{D}_{\dot{\alpha}}\Phi=0$, where 
$\bar{D}_{\dot{\alpha}}=-\bar{\partial}_{\dot{\alpha}}-i(\theta\sigma^\mu)_{\dot{\alpha}}\partial_\mu$
and $\sigma^\mu$ are the four-dimensional sigma matrices.
The supercovariant derivative $\bar{D}_{\dot{\alpha}}$ is used to define
the chiral superfield whilst 
the other supercovariant derivative $D_{\alpha} = \partial_{\alpha} + i
(\sigma^{\mu} \bar{\theta})_{\alpha} \partial_{\mu}$ gives the 
integral measure to form the supersymmetric Lagrangian: 
\begin{eqnarray}
 {\cal L}=-{D^2 \over 4}W(\Phi)+h.c.=\int d^2\theta W(\Phi)+h.c.,
\end{eqnarray}
where $W$ is a superpotential.

Analogously, 
we can construct a superconformal Lagrangian in the
$\mathcal{N} = 3$ projective superspace formalism in three dimensions.
The $\mathcal{N} = 3$ projective superspace consists of the 
ordinary $\mathcal{N} = 3$ superspace $\mathbb{M}^{3|6}$
and the internal space $\mathbb{C}P^1$.
They are parametrized by 
the super-coordinate $z^M = (x^{m},\theta^{\alpha}_I)$ and 
$SU(2)_R$ complex isospinors $v^i, u^i$.
Here $\alpha=1,2$ is the 
$SO(1,2) \sim SL(2,{\bf R})$ spinor and $I = 1,2, 3$ is 
the $SO(3)_R \sim SU(2)_R$ R-symmetry vector 
index respectively. 
We require that the two complex isospinors satisfy the following
completeness relation, 
\begin{eqnarray}
\delta^i {}_j = \frac{1}{(v,u)} (v^i u_j - v_j u^i), \quad
(v,u) \equiv v^i u_i,
\label{completeness}
\end{eqnarray}
where $u_i$ is only restricted by the condition $(v,u) \not= 0$.
We basically use the $SU(2)_R$ spinor indices $i,j = 1,2$ rather than the $SO(3)_R$
vector indices. These are intertwined by the relation
$\theta^{\alpha}_{ij} = (\tau_I)_{ij} \theta^{\alpha}_I$ where
$(\tau_I)^i {}_j$ are the Pauli matrices.
The $SU(2)_R$ indices are raised and
lowered by the anti-symmetric symbols $\varepsilon^{ij}, \varepsilon_{ij}$
such as $\theta^i = \varepsilon^{ij} \theta_j$. 
The basic convention of ordinary superspaces and 
the relation among $\mathcal{N} = 3, 4$ and $\mathcal{N} = 2$
superspaces are presented in Appendix A. 

The supercovariant derivatives in $\mathcal{N} = 3$ superspace are defined by 
\begin{eqnarray}
D^{ij}_{\alpha} = \frac{\partial}{\partial \theta^{\alpha}_{ij}} 
+ i \theta^{\beta}_{ij} \partial_{\alpha \beta},\qquad \partial_{\alpha\beta}\equiv \gamma^m_{\alpha\beta}\partial_m.
\end{eqnarray}
These satisfy the following algebra:
\begin{eqnarray}
\{D^{ij}_{\alpha}, D^{kl}_{\beta} \} = - 2 i \varepsilon^{i(k}
 \varepsilon^{l)j} \partial_{\alpha \beta}.
\end{eqnarray}
Using 
the isospinors $v^i, u^i$ satisfying \eqref{completeness}, 
we define the following set of supercovariant derivatives in the projective superspace:
\begin{eqnarray}
D^{(2)}_{\alpha} = v_i v_j D^{ij}_{\alpha}, \quad
D^{(0)}_{\alpha} = \frac{1}{(v,u)} v_i u_j D^{ij}_{\alpha}, \quad 
D^{(-2)}_{\alpha} = \frac{1}{(v,u)^2} u_i u_j D^{ij}_{\alpha},
\end{eqnarray}
where the superscripts on $D^{(2)}_\alpha$, $D^{(0)}_\alpha$ and
$D^{(-2)}_\alpha$ 
indicate the degree of homogeneity in $v$s.
Among these covariant derivatives, $D^{(2)}_\alpha$ is used to define 
a superconformal projective multiplet. 
We define the superconformal projective multiplet $Q^{(n)}$ with weight 
$n \in \mathbb{Z}$, as a superfield being function of $z^M$ and $v^i$, 
by the following condition,
\begin{eqnarray}
D^{(2)}_{\alpha} Q^{(n)} = 0,
\label{Sconstraint}
\end{eqnarray}
which is an analogy to the chiral condition $\bar{D}_{\dot{\alpha}} \Phi=0$ in
$d=4$, $\mathcal{N}=1$ superfield formalism.
In addition, the superfield $Q^{(n)}$ should be holomorphic with respect to 
$v^i$ and homogeneous function of degree $n$,
\begin{eqnarray}
Q^{(n)} (z, cv) = c^n Q^{(n)} (z, v), \quad c \in \mathbb{C}^{*}.
\label{homogeneous}
\end{eqnarray}
The superconformal transformation of the superfield $Q^{(n)}$ is given by 
\begin{eqnarray}
 \delta Q^{(n)}=-(\xi-\Lambda^{(2)}
\boldsymbol{\partial}^{(-2)})Q^{(n)}-n\Sigma Q^{(n)},
\end{eqnarray}
where $\xi = \xi^m \partial_m + \xi^{\alpha}_I D^I_{\alpha}$ with $D_\alpha^{ij}=(\tau_I)^{ij}D_\alpha^I$
is the superconformal killing vector field and 
$\boldsymbol{\partial}^{(-2)}={1 \over (v,u)}u^i{\partial \over \partial
v^i}$ is the differentiation with respect to the isospinor $v^i$.
The parameters $\Lambda^{(2)}=v_iv_j\Lambda^{ij}$ 
and $\Sigma$ correspond to $SO(3)_R$ and the scale transformations 
respectively. 
Detailed explanation of the superconformal transformation 
is found in \cite{KuPaTaUn}.
A conjugation which is consistent with the constraint
\eqref{Sconstraint} is called the smile conjugation.
This is defined by 
\begin{eqnarray}
\breve{Q}^{(n)} (v) \equiv \left. \overline{Q^{(n)} (v)}  \right|_{\overline{v^i} \to -v_i},
\end{eqnarray}
where the bar stands for the ordinary complex conjugation and
the conjugation of $v^i$ is, more explicitly,
$\overline{v^1} \rightarrow -v_1 = v^2$ and $\overline{v^2} \rightarrow -v_2 = -v^1$.

Now we construct the $\mathcal{N} = 3$ superconformal invariant action.
The supercovariant derivative $D^{(2)}_{\alpha}$ 
has been used to define the projective multiplet
while the others $D^{(0)}_{\alpha}, D^{(-2)}_{\alpha}$ 
are used to form the Grassmann integral measure in an $\mathcal{N}=3$ superconformal action.
The resultant action is given by \cite{KuPaTaUn}:
\begin{eqnarray}
S = 
 \frac{1}{8\pi} \oint_{\gamma} \!
(v, dv) 
 \int \! d^3 x 
\ (D^{(-2)})^2 (D^{(0)})^2 \left. \mathcal{L}^{(2)} (z,v)
 \right|_{\theta = 0},
\label{action}
\end{eqnarray}
where $\mathcal{L}^{(2)}$ is an weight-2 real superconformal projective multiplet.
Note that the action is formed so that sum of the degree of homogeneity in $v$s is zero. 
We sometimes call $\mathcal{L}^{(2)}$ Lagrangian.
The line integral is evaluated over a closed contour $\gamma$ in $\mathbb{C}P^1$. 
Along the contour, $u^i$ should satisfy $(v, u) \not=0$. 
It is shown that the action \eqref{action} is $u$ independent and 
we can therefore choose $u_i = (1,0)$.

In the following, we rewrite the action (\ref{action}) to the one in terms of 
$\mathcal{N}=2$ superspace and superfields.
Without loss of generality, we can take the contour $\gamma$ 
in (\ref{action})  such that it does not pass through the north pole $v^i=(0,1)$ 
in $\mathbb{C}P^1$.
It is then useful to introduce a complex inhomogeneous coordinate 
$\zeta \in \mathbb{C}$ in the upper hemisphere (north chart) of $\mathbb{C}P^1$,
\begin{eqnarray}
v^i = v^1 (1, \zeta), \quad \zeta \equiv \frac{v^2}{v^1}, \quad i=1,2.
\end{eqnarray}
We consider the projective multiplet in this chart.
As we will see, this $\zeta$ is identified with the projective
coordinate in the ordinary projective superspace formalism \cite{KaLiRo}.
Using the coordinate $\zeta$, the covariant derivative
$D^{(2)}_{\alpha}$ turns into the form
\begin{eqnarray}
\begin{aligned}
 & D^{(2)}_{\alpha} = (v^1)^2 D^{[2]}_{\alpha},  \\
 & D^{[2]}_{\alpha} (\zeta) \equiv - \bar{\mathbb{D}}_{\alpha} - 2 \zeta
 D^{12}_{\alpha} + \zeta^2 \mathbb{D}_{\alpha}, \label{cov2}
\end{aligned}
\end{eqnarray}
where we have introduced the $\mathcal{N} = 2$ supercovariant derivatives
$\mathbb{D}_{\alpha}$ and $\bar{\mathbb{D}}_{\alpha}$ (see Appendix A).
As discussed in \cite{KuPaTaUn}, all the $v^1$ dependence of the
superconformal projective multiplet $Q^{(n)} (z,v)$ can be factored out and 
a new superfield $Q^{[n]} (z,v)\propto Q^{(n)} (z,v)$ is defined.
With the use of this fact and (\ref{cov2}), we find that the constraint (\ref{Sconstraint}) becomes
\begin{eqnarray}
D^{[2]}_{\alpha} (\zeta) Q^{[n]} (z, \zeta) = 0. 
\label{Pconstraint}
\end{eqnarray}
In general, $Q^{[n]}$ is expanded by power series 
in $\zeta$,
\begin{eqnarray}
Q^{[n]} (z, \zeta) = \sum_k \zeta^k Q_k (z),
\end{eqnarray}
where $Q_k (z)$ are some ordinary $\mathcal{N} = 3$ superfields subject
to the constraints \eqref{Pconstraint}.
Using the factorization of $v^1$, the projective multiplet $\mathcal{L}^{(2)}$ is rewritten as
\begin{eqnarray}
\mathcal{L}^{(2)} (z, v) = (v^1)^2 (i\zeta) \mathcal{L}^{[2]} (z, \zeta),
\end{eqnarray}
and the action \eqref{action} 
reduces to the following form,
\begin{eqnarray}
S = \frac{1}{2\pi i} \oint_{\gamma} \! \frac{d\zeta}{\zeta} 
\int d^3 x d^4 \theta \left. \mathcal{L}^{[2]} (z, \zeta)
\right|_{\theta_{12} = 0},
\label{N2action}
\end{eqnarray}
where we have used (\ref{cov2}) and (\ref{Pconstraint}).
The factor $i\zeta$ also appears in addition to $(v^1)^2$ for reality of 
$\mathcal{L}^{(2)}$ (see also $\mathcal{O}(-k,k)$ and tropical
multiplets below).
The expression \eqref{N2action} is analogous to the action in $d = 4$,
$\mathcal{N} = 2$ ordinary projective superspace 
and is completely determined by 
the $\mathcal{N} = 3$ superfields $Q_k (z)$ projected on the $\mathcal{N} =
2$ superspace.
Note that even though the integration in the action is carried out over the $\mathcal{N}
= 2$ superspace only, the action has $\mathcal{N} = 3$ off-shell supersymmetry by construction. 

We now give several examples of the projective superfields $Q^{(n)}$. 
\\
\\
\underline{\textbullet \ $\mathcal{O}(k)$ and (ant)arctic multiplets}
\\
\\
The weight-$n$ complex $\mathcal{O}(k)$ multiplet is defined to be holomorphic in the
north chart of $\mathbb{C}P^1$,
\begin{eqnarray}
\Upsilon^{(n)} (z,v) = (v^1)^n \Upsilon^{[n]} (z, \zeta), \quad 
\Upsilon^{[n]} = \sum^{k}_{l=0} \zeta^l \Upsilon_l (z).
\label{Ok_expansion}
\end{eqnarray}
The constraints \eqref{Pconstraint} on the $\mathcal{N} = 3$ component
superfields are given by 
\begin{eqnarray}
\begin{aligned}
 & \bar{\mathbb{D}}_{\alpha} \Upsilon_0 = 0, \\
 & \bar{\mathbb{D}}_{\alpha} \Upsilon_1 + 2 D^{12}_{\alpha} \Upsilon_0
 = 0, \\
 & \bar{\mathbb{D}}_{\alpha} \Upsilon_l + 2 D_{\alpha}^{12}
 \Upsilon_{l-1} - \mathbb{D}_{\alpha} \Upsilon_{l-2} = 0, \quad 
(2 \le l \le k), \\
 &  2 D^{12}_{\alpha} \Upsilon_k - \mathbb{D}_{\alpha} \Upsilon_{k-1} =
 0, \\
 & \mathbb{D}_{\alpha} \Upsilon_k = 0.
\end{aligned}
\label{Ok_constraint}
\end{eqnarray}
The arctic multiplet is defined as the limit $k \to \infty$ of the 
complex $\mathcal{O}(k)$ multiplet and its smile conjugate is called the antarctic multiplet.
\\
\\
\underline{\textbullet \ $\mathcal{O}(-k,k)$ and tropical multiplets}
\\
\\
The real $\mathcal{O}(-k,k)$ multiplet with weight $n$ is defined as 
\begin{eqnarray}
\begin{aligned}
&  U^{(2n)} (z,v) = (i v^1 v^2)^n U^{[2n]} (z, \zeta) = (v^1)^{2n} (i
 \zeta)^n U^{[2n]} (z, \zeta),  \\
&  U^{[2n]} (z, \zeta) = \sum^{k}_{l=-k} \zeta^l U_l (z),
 \quad 
\bar{U}_l = (-1)^l U_{-l}.\label{Okk_expansion}
\end{aligned}
\end{eqnarray}
The constraints \eqref{Pconstraint} on the $\mathcal{N} = 3$ component
superfields are given by 
\begin{eqnarray}
\begin{aligned}
 & \bar{\mathbb{D}}_{\alpha} U_{-k} = 0, \\
 & \bar{\mathbb{D}}_{\alpha} U_{- k + 1} + 2 D_{\alpha}^{12} U_{-k} = 0, \\
 & \bar{\mathbb{D}}_{\alpha} U_l + 2 D^{12}_{\alpha} U_{l-1} -
 \mathbb{D}_{\alpha} U_{l-2} = 0, \quad 
 (-k+2 \le l \le k), \\
 & 2 D^{12}_{\alpha} U_k - \mathbb{D}_{\alpha} U_{k-1} = 0, \\
 & \mathbb{D}_{\alpha} U_k = 0.
\end{aligned}
\label{Okk_constraint}
\end{eqnarray}
The tropical multiplet is defined as the limit $k \to \infty$ of the 
$\mathcal{O}(-k,k)$ multiplet. 

\section{$\mathcal{N} = 3$ superconformal Chern-Simons-matter models}
In this section, we construct $\mathcal{N} = 3$ supersymmetric 
Chern-Simons-matter actions with Abelian gauge group in the projective superspace.
\subsection{Chern-Simons term}
We expect that the supersymmetric Chern-Simons 
term is naively given by the product of a  (super)gauge field strength and a gauge potential,
which is inspired by the result in five-dimensional projective superspace \cite{KuLi,Ku2}.
The gauge potential is 
given by the real tropical multiplet $\mathcal{V}^{(0)}(z,v)$ with
weight 
0. 
For the gauge field strength, we assume that it is given by a real $\mathcal{O}(-1,1)$ multiplet 
$G^{(2)}(z,v)$ with weight 
2. 
Then we propose that 
the weight-2 superconformal projective superfield in the action 
is given by 
\begin{eqnarray}
\mathcal{L}^{(2)} = {k \over 8\pi} \mathcal{V}^{(0)} (z,v) G^{(2)} (z,v).
\label{CSLagrangian}
\end{eqnarray} 
Since the tropical multiplet is weight 
0, there is no overall $v^1$ 
dependence, $\mathcal{V}^{(0)} (z,v) = \mathcal{V}^{[0]} (z, \zeta)$.
It is expanded with respect to $\zeta$ as
\begin{eqnarray}
{\cal V}^{[0]}(z,\zeta)=\sum_{n=-\infty}^\infty \zeta^n V_n(\zeta).
\end{eqnarray}
The tropical multiplet transforms under the $U(1)$ gauge transformation: 
\begin{eqnarray}
\delta \mathcal{V}^{[0]} = i (\bar{\Lambda}^{[0]} - \Lambda^{[0]}),\quad
\Lambda^{[0]}=\sum_{n=0}^\infty \zeta^n \lambda_n,\quad
\bar{\Lambda}^{[0]}=\sum_{n=0}^\infty (-1/\zeta)^n\bar{\lambda}_n,
\label{gauge_transf}
\end{eqnarray}
where $\Lambda^{[0]}(z,\zeta)$ and $\bar{\Lambda}^{[0]}(z,\zeta)$ are gauge parameters in the 
weight-0 arctic and antarctic multiplets $\Lambda^{(0)}(z,\zeta)$ and 
$\bar{\Lambda}^{(0)}(z,\zeta)$ respectively. 
The gauge transformations of the real tropical multiplet in 
terms of $\zeta$-expansion coefficients are 
\bea
 \delta V_0 = i(\bar{\lambda}_0- \lambda_0),\quad 
 \delta V_n = - i\lambda_n,
     \label {a-trans}
\eea
By the gauge transformation, we can 
make the tropical multiplet be a real $\mathcal{O} (-1,1)$ multiplet form 
(Lindstr{\"o}m-Ro\v{c}ek gauge) \cite{LiRo}, 
\begin{eqnarray}
\mathcal{V}^{[0]} = \frac{1}{\zeta} V_{-1} + V_0 + \zeta V_1.
\end{eqnarray}
The weight-2 $\mathcal{O}(-1,1)$ multiplet becomes 
$G^{(2)} = (v^1)^2 (i\zeta) G^{[2]}(z,\zeta)$
where $G^{[2]}$ is expanded as 
\begin{eqnarray}
 G^{[2]} (z, \zeta) = 
{i \over \zeta} \Phi + L + i \zeta \bar{\Phi}. \label{o2}
\end{eqnarray}
We assume that $G^{[2]}$ is invariant under 
the $U(1)$ gauge transformation. Indeed, as
we will see in below, the components in $G^{[2]}$ are written in terms of
the components in 
$\mathcal{V}^{[0]}$  as the gauge invariant forms.

As we have discussed in the previous section, 
the constraint \eqref{Pconstraint} is interpreted as the constraints on 
the $\mathcal{N} = 3$ component superfields. 
Since the action is totally expressed in terms of 
$\mathcal{N} = 3$ component superfields
projected on the $\mathcal{N} = 2$ superspace, 
we are interested in the constraints in the $\mathcal{N} = 2$ subsuperspace.
From the constraints \eqref{Ok_constraint} and 
\eqref{Okk_constraint}, we can find the constraints for the 
component superfields.
For $\mathcal{V}^{(0)}$, since this is the tropical multiplet, $V_0|$ is a
real superfield and other fields are unconstrained.
Here the symbol ``$|$'' means the 
$\mathcal{N}=2$ projection of $\mathcal{N}=3$ superfield.
For the $\mathcal{O}(-1,1)$ multiplet $G^{[2]}$, we have constraints 
\begin{eqnarray}
\begin{aligned}
 & \bar{\mathbb{D}}_{\alpha} \Phi| = 0, \\
 & \bar{\mathbb{D}}^2 L| = \mathbb{D}^2 L| = 0, \\
 & \mathbb{D}_{\alpha} \bar{\Phi}| = 0.
\end{aligned}
\label{Gconst}
\end{eqnarray}
These conditions imply that $\Phi|$ and $\bar{\Phi}|$ are the $\mathcal{N} =
2$ chiral and antichiral superfields while $L|$ is the real linear superfield.
Finally, for the gauge parameters 
$\Lambda^{[0]} (z,\zeta)$ and $\bar{\Lambda}^{[0]} (z,\zeta)$, we have
\begin{eqnarray}
\begin{aligned}
 & \bar{\mathbb{D}}_{\alpha} \lambda_0| = \mathbb{D}_{\alpha}
 \bar{\lambda}_0| = 0,  \\
 & \bar{\mathbb{D}}^2 \lambda_1| = \mathbb{D}^2 \bar{\lambda}_1| = 0,
\end{aligned}
 \label{Lconst}
\end{eqnarray}
and other fields are unconstrained. Therefore $\lambda_0|$ and
$\bar{\lambda}_0|$ are $\mathcal{N} = 2$ (anti)chiral while $\lambda_1|$ and
$\bar{\lambda}_1|$ are $\mathcal{N} = 2$ (anti)linear superfields. 
In the following we omit the symbol ``$|$'' and consider 
the component superfields in $\mathcal{N} = 2$ superspace.

After fixing $u$ to $u_i = (1,0)$, the action \eqref{action} with 
the Lagrangian (\ref{CSLagrangian}) reduces to the following form,
\begin{eqnarray}
S_{\rm CS} = {k \over 8\pi} \oint_\gamma {d\zeta \over 2\pi i\zeta} 
\int \! d^3x d^4 \theta \ {\cal V}^{[0]} (z,\z) G^{[2]}(z,\zeta). 
\label{a-CS}
\end{eqnarray}
The component superfields $L$, $\Phi$ and $\bar{\Phi}$ should be expressed 
in terms of $V_n$ and they satisfy the constraints 
(\ref{Gconst}). 
We find that 
the solutions to the constraints are given by 
\bea
 L=i\bar{\mathbb{D}}^\a \mathbb{D}_\a V_0, \quad 
\Phi = {i \over 8}\bar{\mathbb{D}}^\a \bar{\mathbb{D}}_\a V_1,\quad 
\bar{\Phi} = 
-
{i \over 8} \mathbb{D}^\a \mathbb{D}_\a V_{-1}. \label{o2-comp}
\eea
One can confirm that all these expressions are gauge invariant by using 
the relations \eqref{Lconst}.
This fact is consistent with our assumption.
The solutions \eqref{o2-comp} are also expressed as the integral of the
tropical multiplet over the isospinor space:
\begin{eqnarray}
G^{(2)} (w)&=& - \frac{1}{8 \pi i} \oint_{\gamma} \! (v, dv)
\left\{
- \frac{i}{2} (w,v)^2 D^{(-2) \alpha} D^{(-2)}_{\alpha}
\right. 
\nonumber \\
& & \qquad 
\left. 
+ 8 \frac{(w,v)(w,u)}{(v,u)} D^{(-2) \alpha} D^{(0)}_{\alpha}   
+ 2 i \frac{(w,u)^2}{(v,u)^2} D^{(0)\alpha} D^{(0)}_{\alpha}
\right\} \mathcal{V}^{(0)} (v),
\label{N3sol}
\end{eqnarray}
This closed form expression is obtained 
in the context of three-dimensional supergravity in the projective
superspace \cite{KuLiTa}
\footnote{We have slightly changed the coefficients to make it be
consistent with our convention.}.
After fixing $u = w$ and using the completeness relation of the
isospinor, we find that \eqref{N3sol} is also expressed as
\begin{eqnarray}
G^{(2)} (w) = - \frac{1}{8 \pi i} (D^{(2)} (w))^2 
\oint_{\gamma} \frac{(v,dv)}{(v,w)} \mathcal{V}^{(0)} (v).
\end{eqnarray}
The detailed calculations are found in Appendix B.

Next, we examine the gauge invariance of the action.
Since the action \eqref{action} is completely equivalent to the
expression \eqref{N2action} and the gauge variation $\delta
\mathcal{V}^{[0]}$ is the tropical 
multiplet with weight 0, it is sufficient to show the gauge
invariance of \eqref{a-CS} rather than 
its manifest $\mathcal{N} = 3$ action \eqref{action}.
Substituting the expressions (\ref{a-trans}) and (\ref{o2})  
into 
the gauge variation of the action $\delta S_{\mathrm{CS}}$
and integrating it with respect to $\z$, we find 
\begin{eqnarray}
 \delta S_{\rm CS}={k \over 8\pi}\int d^3 x d^4\theta[\bar{\lambda}_{1} \bar{\Phi}
  +i( \bar{\lambda}_0 - \lambda_0 )L + \lambda_1 \Phi].
\end{eqnarray}
This expression is vanishing, taking account of the constraints 
(\ref{Gconst}), (\ref{Lconst}) and the relations  
$d^2\theta=-\mathbb{D}^\alpha \mathbb{D}_\alpha/4$, 
$d^2\bar{\theta}=-\bar{\mathbb{D}}^\alpha \bar{\mathbb{D}}_\alpha/4$
in the space-time integration.
Therefore $S_{\rm CS}$ is gauge invariant.
Note that the higher order components in $\zeta^n$ are dropped in the
expression due to the $\zeta$-integration.

To see the action in terms of $\mathcal{N} =2$ superfields, 
we substitute the solution (\ref{o2-comp}) into the action (\ref{a-CS}) and find
\begin{eqnarray}
 S_{\rm CS} &=&{k \over 8\pi}\int d^3 x d^4 \theta 
   (i V_1\Phi + i V_0\bar{\mathbb{D}}^\a
   \mathbb{D}_\a V_0 
+ i V_{-1} \bar{\Phi}) \non \\
 &=&{ik \over 8\pi}\int d^3 x d^4\theta \ V_0 \bar{\mathbb{D}}^\a
  \mathbb{D}_\a V_0 \non \\
 && \quad + {i k \over 8\pi}\int d^3 x d^2\theta 
\left(-{\bar{\mathbb{D}}^\alpha\bar{\mathbb{D}}_\alpha \over 4}V_{1} \right)\Phi
   + {i k \over 8\pi}\int d^3 x d^2\bar{\theta}
   \left(-{\mathbb{D}^\alpha\mathbb{D}_\alpha \over 4}
V_{-1} \right)\bar{\Phi} \non \\
 &=&{ik \over 8\pi}\int d^3 x d^4\theta \ V_0 \bar{\mathbb{D}}^\a \mathbb{D}_\a V_0 
   - {k \over 4\pi}\int d^3 x d^2\theta \ \Phi^2
+ {k \over 4\pi}\int d^3 x d^2\bar{\theta} \ \bar{\Phi}^2.
\label{N3CSN2}
\end{eqnarray}
This is equivalent to the expression given in (\ref{N2CSHU1}).
Therefore the Lagrangian \eqref{CSLagrangian} correctly
reproduces the $\mathcal{N} = 3$ Chern-Simons action in 
$\mathcal{N} = 2$ superfield formalism.

\subsection{Matter part}
It is known that the matter hypermultiplet is embedded into the 
ant(arctic) multiplet. We consider the weight-1 arctic multiplet $\Upsilon^{(1)}$ 
as the matter part of the $\mathcal{N} = 3$ Chern-Simons-matter model. 
The Lagrangian of the free (ant)arctic multiplets is given by 
\begin{eqnarray}
\mathcal{L}^{(2)} = i\breve{\Upsilon}^{(1)} \Upsilon^{(1)} 
= i(v^1)^2 \zeta \bar{\Upsilon}^{[1]} \Upsilon^{[1]}.
\end{eqnarray}
This is easily expanded into the $\cN=3$ component superfields.
Substituting the expansion \eqref{Ok_expansion} into the action
\eqref{N2action} and performing the $\zeta$-integration, 
we find
\bea
 S_{\rm free} 
= \int d^3 x d^4\theta \sum_{l=0}^\infty (-1)^l \bar{\U}_l \U_l.
\eea
According to the constraints (\ref{Ok_constraint})
on the $\mathcal{N} = 2$ projected superfields, 
only $\U_0$ ($\bar{\Upsilon}_0$) and $\U_1$ ($\bar{\Upsilon}_1$) are dynamical fields
and the others are auxiliary fields, which are integrated out with their equations of motion.
The resulting action describes 
the four-dimensional $\mathcal{N} = 2$ massless free hypermultiplet 
dimensionally reduced to three dimensions.

We now introduce a gauge interaction.
We consider the action of the (ant)arctic multiplets coupled to the Abelian gauge field.
Following the ordinary projective superfield approach in four dimensions 
\cite{GoUn}, the Lagrangian is given by
\begin{eqnarray}
\mathcal{L}^{(2)} = i \breve{\Upsilon}^{(1)} \exp(\mathcal{V}^{(0)})
 \Upsilon^{(1)} = i(v^1)^2 \zeta \bar{\Upsilon}^{[1]}
 \exp(\mathcal{V}^{[0]}) \Upsilon^{[1]}.
\end{eqnarray}
The action \eqref{N2action} 
for this Lagrangian is given by 
\bea
 S_{\rm matter} = \int d^3x d^4\theta \oint_\gamma {d \z \over 2 \pi i \z} \bar{\U}^{[1]}
 e^{\mathcal{V}^{[0]}} \U^{[1]}.
\label{higgs}
\eea
It is easy to confirm that the action is 
invariant under the following $U(1)$ gauge transformations
\bea
 &\U^{[1]\prime} = e^{i \Lambda^{[0]} }\U^{[1]},\quad
 \bar{\U}^{[1]\prime} = e^{-i\bar{\Lambda}^{[0]}}\bar{\U}^{[1]},\quad 
  (e^{\cV^{[0]}})^{\prime} = e^{i\bar{\Lambda}^{[0]}}e^{\cV^{[0]}} e^{-i\Lambda^{[0]}}.&
\eea
We write down the action \eqref{higgs} in terms of $\mathcal{N} = 2$
superfields. 
In the Abelian case, it is always possible to split the tropical multiplet $\mathcal{V}^{[0]}$ into
the arctic $\mathcal{V}_{+}$ and the antarctic $\mathcal{V}_{-}$ pieces,
\begin{eqnarray}
e^{\mathcal{V}^{[0]}} = e^{\mathcal{V}_{+}} e^{\mathcal{V}_{-}},
\end{eqnarray}
where we have defined
\begin{eqnarray}
\mathcal{V}_{+} = \frac{1}{2} V_0 + \sum^{\infty}_{n=1} \zeta^n
 V_n, \quad \mathcal{V}_{-} = \frac{1}{2} V_0 + \sum_{n= 1}^{\infty} \zeta^{-n} V_{-n}.
\end{eqnarray}
In the following, we employ the Lindstr{\"o}m-Ro\v{c}ek gauge so that the tropical
multiplet $\mathcal{V}^{[0]}$ takes the $\mathcal{O} (-1,1)$ multiplet
form. We define the gauge supercovariant derivative as 
\begin{eqnarray}
\mathcal{D}_{\alpha}^{[2]} &=& e^{- q \mathcal{V}_{-}} D_{\alpha}^{[2]}
 e^{q \mathcal{V}_{-}} = e^{q \mathcal{V}_{+}} D_{\alpha}^{[2]} e^{- q
 \mathcal{V}_{+}}
\nonumber \\
&\equiv& - \bar{\mathcal{D}}_{\alpha} - 2 \zeta \mathcal{D}^{12}_{\alpha} +
 \zeta^2 \mathcal{D}_{\alpha}, \label{cd}
\end{eqnarray}
where $q$ is the $U(1)$ charge of associated fields and 
\begin{eqnarray}
\begin{aligned}
\mathcal{D}_{\alpha} =& \mathbb{D}_{\alpha} + \frac{q}{2}
 \mathbb{D}_{\alpha} 
V_0, \\
\mathcal{D}^{12}_{\alpha} =& D_{\alpha}^{12} - \frac{q}{2} 
\left(
\mathbb{D}_{\alpha} V_{-1} - D^{12}_{\alpha} V_0
\right), \\
\bar{\mathcal{D}}_{\alpha} =& \bar{\mathbb{D}}_{\alpha} + \frac{q}{2} 
\left(
\bar{\mathbb{D}}_{\alpha} 
V_0 + 4 D^{12}_{\alpha} V_{-1}
\right).
\end{aligned}
\end{eqnarray}
The second equality in the first line in (\ref{cd}) holds because of 
$D_\alpha^{[2]}e^{\mathcal{V}^{[0]}}=0$.
The anticommutation relations of these supercovariant derivatives are found
in Appendix C.
The action \eqref{higgs} is rewritten as 
\begin{eqnarray}
S_{\rm matter} = \int \! d^3 x d^4 \theta  \oint_{\gamma} \frac{d \zeta}{2 \pi i \zeta} 
\tilde{\bar{\Upsilon}}^{[1]} \tilde{\Upsilon}^{[1]},
\label{matter_action}
\end{eqnarray}
where we have defined 
\begin{eqnarray}
\begin{aligned}
\tilde{\Upsilon}^{[1]} =& e^{\mathcal{V}_{+}} \Upsilon^{[1]} =
 \sum_{n=0}^{\infty} \zeta^n \tilde{\Upsilon}_n, \\
\tilde{\bar{\Upsilon}}^{[1]} =& e^{\mathcal{V}_{-}} \bar{\Upsilon}^{[1]}
= \sum^{\infty}_{n=0} \left(-1\over \zeta\right)^{n}
 \tilde{\bar{\Upsilon}}_{n}. 
\end{aligned}
\end{eqnarray}
We can show that these superfields satisfy the gauge covariantized
projective constraints 
$\mathcal{D}^{[2]}_{\alpha} \tilde{\Upsilon}^{[1]} =
\mathcal{D}^{[2]}_{\alpha} \tilde{\bar{\Upsilon}}^{[1]} = 0$
and give the gauge covariant (ant)arctic multiplets.
The constraints on the component superfields are found to be 
\begin{eqnarray}
&& \bar{\mathcal{D}}_{\alpha} \tilde{\Upsilon}_0 = 0, \label{const0}\\
&& \bar{\mathcal{D}}_{\alpha} \tilde{\Upsilon}_1+2\mathcal{D}^{12}_\alpha\tilde{\U}_0 = 0, \label{const1} \\
&& \bar{\mathcal{D}}_{\alpha} \tilde{\Upsilon}_l + 2
 \mathcal{D}^{12}_{\alpha} \tilde{\Upsilon}_{l-1} - \mathcal{D}_{\alpha}
 \tilde{\Upsilon}_{l-2} = 0, \quad (l \ge 2), \\
&& \mathcal{D}_{\alpha} \tilde{\bar{\Upsilon}}_{0} = 0, \label{constb0}\\
&& \mathcal{D}_\alpha \tilde{\bar{\U}}_1-2\mathcal{D}_\alpha^{12}\tilde{\bar{\U}}_0=0, \label{const2}\\
&& \mathcal{D}_{\alpha} \tilde{\bar{\Upsilon}}_{l-2} + 2
 \mathcal{D}^{12}_{\alpha} \tilde{\bar{\Upsilon}}_{l-1} -
 \mathcal{D}_{\alpha} \tilde{\bar{\Upsilon}}_l = 0, \quad (l \ge 2).
\end{eqnarray}
With the use of the commutation relations (\ref{B1}) and (\ref{Okk_constraint}), the constraints (\ref{const1}) and (\ref{const2}) turn out to be
\begin{eqnarray}
 \bar{\mathcal{D}}^2\tilde{\U}_1= 
\bar{\mathbb{D}}^2V_1\tilde{\U}_0,\quad
 \mathcal{D}^2\tilde{\bar{\U}}_1 = 
-
{\mathbb{D}}^2V_{-1}\tilde{\bar{\U}}_0.
\end{eqnarray}
The matter action \eqref{matter_action} is now expanded as 
\begin{equation}
S_{\mathrm{matter}} = \int \! d^3 x d^4 \theta \
\left[
\tilde{\bar{\Upsilon}}_0 \tilde{\Upsilon}_0 - \tilde{\bar{\Upsilon}}_1
\tilde{\Upsilon}_1 
+ \tilde{Y} 
\left(
\mathcal{D}^2 \tilde{\bar{\Upsilon}}_1 
+ 
\mathbb{D}^2 V_{-1} \tilde{\bar{\Upsilon}}_0
\right)
+
\tilde{\bar{Y}}
\left(
\bar{\mathcal{D}}^2 \tilde{\Upsilon}_1 
- 
\bar{\mathbb{D}}^2 V_1 \tilde{\Upsilon}_0
\right)
\right],
\end{equation}
where we have integrated out the auxiliary fields $\tilde{\Upsilon}_{l},
\tilde{\bar{\Upsilon}}_{l}$ $(l \ge 2)$.
We have also introduced the Lagrange multipliers $\tilde{Y},
\tilde{\bar{Y}}$ which are related to the non-covariantized Lagrange
multipliers $Y$ and $\bar{Y}$ through the relation 
$\tilde{Y} = e^{- \frac{1}{2} V_0} Y$, $\tilde{\bar{Y}} = e^{-
\frac{1}{2} V_0} \bar{Y}$.
Their gauge transformations are given by $Y \to e^{i \bar{\lambda}_0} Y$,
$\bar{Y} \to e^{- i \lambda_0} \bar{Y}$.
After integrating out $\tilde{\bar{\Upsilon}}_1, \tilde{\Upsilon}_1$ and 
using the constraints 
(\ref{const0}), (\ref{constb0}), the solutions \eqref{o2-comp}, 
and the anticommutation relations of the supercovariant derivatives 
\eqref{B1}, the action becomes 
\begin{equation}
S_{\mathrm{matter}} = \int \! d^3 x d^4 \theta \
\left[
\bar{S} e^{V_0} 
S + 
T e^{- V_0} 
\bar{T} 
\right]
+ 2 \int \! d^3 x d^2 \theta \ 
T \Phi S 
- 2 \int \! d^3 x d^2 \bar{\theta} \ 
\bar{T} \bar{\Phi} \bar{S}.
\label{N3matter}
\end{equation}
Here we have relabeled 
$S = \Upsilon_0$, $\bar{S} = \bar{\Upsilon}_0$ 
and have defined 
\begin{equation}
\tilde{\bar{T}} = e^{- \frac{1}{2} V_0} \bar{T} = i \mathcal{D}^2 \tilde{Y}, \quad
\tilde{T} = e^{- \frac{1}{2} V_0} T = 
 i \bar{\mathcal{D}}^2 \tilde{\bar{Y}}.
\end{equation}
The gauge transformations of $\bar{T}, T$ are the same as the ones for
the Lagrange multipliers $Y, \bar{Y}$. 
The action \eqref{N3matter} agrees with \eqref{N3higgs} with $N_f = 1$. 
Generalization to the arbitrary number of flavours $N_f$ is straightforward.

\section{$\mathcal{N} = 4$ superconformal Chern-Simons theories}
In this section, we generalize 
the $\mathcal{N} = 3$ construction discussed in the previous section 
to $\mathcal{N} = 4$ cases. 
The $\mathcal{N} = 4$ superspace $\mathbb{M}^{3|8}$ is parametrized by
the super-coordinates
$
z^M = (x^m, \theta^{\alpha}_{i \bar{j}}),
$
where $i = 1,2$ and $\bar{j} = 1,2$ are indices for the 
$SU(2)_L \times SU(2)_R$ subgroup of $SO(4)_R$.
These indices are raised and lowered by the antisymmetric matrices
$\varepsilon^{ij}, \varepsilon^{\bar{i} \bar{j}}$ and so on. 
To incorporate the two $SU(2)$ symmetries, 
we introduce a pair of $\mathbb{C}P^1$, namely, 
the mirror projective spaces $\mathbb{C}P^1_L \times \mathbb{C}P^1_R$ \cite{KuPaTaUn}.
As a result, the $\mathcal{N} = 4$ projective superspace is given by 
$\mathbb{M}^{3|8} \times \mathbb{C}P^1_L \times \mathbb{C}P^1_R$. 
The mirror projective spaces $\mathbb{C}P^1_L \times \mathbb{C}P^1_R$ 
are parametrized by the homogeneous complex coordinates $v_L = (v^i), v_R =
(v^{\bar{i}})$. As in the case of $\mathcal{N} = 3$, 
they are supplemented by $u_i, u_{\bar{k}}$ and satisfy the 
similar completeness relations as in \eqref{completeness}.

The supercovariant derivatives 
in $\mathcal{N} = 4$ superspace 
are given by 
\begin{eqnarray}
D^{i \bar{j}}_{\alpha} = \frac{\partial}{\partial \theta^{\alpha}_{i
 \bar{j}}} + i \theta^{\beta}_{i \bar{j}} \partial_{\alpha \beta}.
\end{eqnarray}
They satisfy the following algebra
\begin{eqnarray}
\{D^{i \bar{j}}_{\alpha}, D^{k \bar{l}}_{\beta} \} = 
2 i \varepsilon^{ik} \varepsilon^{\bar{j} \bar{l}} \partial_{\alpha \beta}.
\end{eqnarray}
As in the $\mathcal{N} = 3$ case, we define the following
covariant derivatives:
\begin{eqnarray}
D^{(1) \bar{k}}_{\alpha} = v_i D^{i \bar{k}}_{\alpha}, \quad
D^{(-1) \bar{k}}_{\alpha} = \frac{1}{(v_L, u_L)} u_i D^{i\bar{k}}_{\alpha},
\end{eqnarray}
and 
\begin{eqnarray}
D^{(1)i}_{\alpha} = v_{\bar{k}} D^{i \bar{k}}_{\alpha}, \quad 
D^{(-1) i}_{\alpha} = \frac{1}{(v_R, u_R)} u_{\bar{k}} 
D^{i \bar{k}}_{\alpha}.
\end{eqnarray}
The covariant derivatives $D^{(1){\bar{k}}}_{\alpha},
D^{(-1)\bar{k}}_{\alpha}$ satisfy the algebras,
\begin{eqnarray}
\begin{aligned}
 & \{ D^{(1) \bar{k}}_{\alpha}, D^{(1)\bar{l}}_{\beta} \} = 
\{ D^{(-1)\bar{k}}_{\alpha}, D^{(-1)\bar{l}}_{\beta} \} = 0, \\
 & \{D^{(1) \bar{k}}_{\alpha}, D^{(-1) \bar{l}}_{\beta}\} = - 2 i
 \varepsilon^{\bar{k} \bar{l}} \partial_{\alpha \beta}. 
\end{aligned}
\end{eqnarray}
The other covariant derivatives $D^{(1)i}_{\alpha}, D^{(-1)i}_{\alpha}$ 
satisfy the similar algebras. 
In the $\mathcal{N} = 4$ case, we can introduce the left and right
weight-$n$ projective
multiplets associated with each $\mathbb{C}P^1$.
They are defined by
\begin{eqnarray}
\begin{aligned}
 & D^{(1) \bar{k}}_{\alpha} Q^{(n)}_L (v_L) = 0, \\
 & D^{(1) i}_{\alpha} Q^{(n)}_R (v_R) = 0. \label{constN4}
\end{aligned}
\end{eqnarray}
Each projective multiplet $Q^{(n)}_L$, $Q^{(n)}_R$ satisfies 
the relation \eqref{homogeneous}.
The $\mathcal{N} = 4$ superconformal transformations of the left and
right projective superfields are given by 
\begin{eqnarray}
\begin{aligned}
\delta Q^{(n)}_L =& - \left(
\xi - \Lambda^{(2)}_L 
\boldsymbol{\partial}_L^{(-2)} 
\right) Q^{(n)}_L - n \Sigma_L Q^{(n)}_L, 
\\
\delta Q^{(n)}_R =&  - \left( 
\xi - \Lambda^{(2)}_R 
\boldsymbol{\partial}_R^{(-2)} 
\right) Q^{(n)}_R - n \Sigma_R Q^{(n)}_R,
\end{aligned}
\end{eqnarray}
where $\xi$ is the superconformal Killing vector field, $\Lambda_{L,R}$, 
$\Sigma_{L,R}$ and $\boldsymbol{\partial}^{(-2)}_{L,R}$ 
are defined as in the same way in $\mathcal{N} = 3$ case. 
See \cite{KuPaTaUn} for details.

Following the previous section, we introduce the
complex inhomogeneous coordinate $\zeta_L$ in the left part 
as 
\begin{eqnarray}
v^i = v^1 (1, \zeta_L), \quad \zeta_L = \frac{v^2}{v^1}.
\end{eqnarray}
The covariant derivative becomes
\begin{eqnarray}
D^{(1)\bar{k}}_{\alpha} = v^1 D^{[1]\bar{k}}_{\alpha}, \qquad 
D^{[1]\bar{k}}_{\alpha} \equiv D_{\alpha}^{2\bar{k}} - \zeta D^{1 \bar{k}}_{\alpha}.
\end{eqnarray}
The $v^1$ dependencies of the projective superfields 
can be factored out and one can define 
$Q^{[n]}_L \propto Q^{(n)}_L$ that 
satisfies
\begin{equation}
D^{[1]\bar{k}}_{\alpha} (\zeta) Q^{[n]}_L = 0.
\label{N4Lconstraint}
\end{equation}
Then, the left projective superfield $Q^{[n]}_L$ is expanded as 
\begin{eqnarray}
Q^{[n]}_L (z, \zeta_L) = \sum_k \zeta_L^k Q_k (z),
\end{eqnarray}
where $Q_k(z)$ are $\mathcal{N} = 4$ superfields
subject to the constraint (\ref{constN4}). 
Similar definitions hold in the right part.

The $\mathcal{N} = 4$ superconformal invariant action 
is given by 
\begin{eqnarray}
S = 
\frac{1}{2\pi} \oint_{\gamma_L} \! 
(v_L, dv_L)
\int \! d^3 x \ D^{(-4)}_L \mathcal{L}_L^{(2)} (z, v_L) |_{\theta = 0} 
+
\frac{1}{2\pi} \oint_{\gamma_R} \! 
(v_R, dv_R)
\int \! d^3 x \ D^{(-4)}_R \mathcal{L}_R^{(2)} (z, v_R) |_{\theta = 0},
\nonumber \\
\label{N4action}
\end{eqnarray}
where $\mathcal{L}^{(2)}_L(\mathcal{L}^{(2)}_R)$ is the weight-2 left (right) projective multiplet and
we have defined
\begin{eqnarray}
\begin{aligned}
 & D^{(-4)}_L = \frac{1}{48} D^{(-2) \bar{k} \bar{l}} D_{\bar{k}
 \bar{l}}^{(-2)}, \quad D^{(-2)}_{\bar{k} \bar{l}} = D^{(-1)
 \alpha}_{\bar{k}} 
D^{(-1)}_{\alpha \bar{l}}, \\
 & D^{(-4)}_R = \frac{1}{48} D^{(-2)ij} D_{ij}^{(-2)}, \quad
 D^{(-2)}_{ij} = D^{(-1)
 \alpha}_i D^{(-1)}_{\alpha j}.
\end{aligned}
\end{eqnarray}
The contour $\gamma_L$ $(\gamma_R)$ is chosen such that the path goes
the outside of the north pole in $\mathbb{C}P^1_L$ $(\mathbb{C}P^1_R)$.
After fixing $u_i = (1,0)$, $u_{\bar{k}} = (1,0)$ in $\mathbb{C}P^1_L$
and $\mathbb{C}P^1_R$, the action can be
rewritten as the one in $\mathcal{N} = 2$ superspace:
\begin{eqnarray}
S = 
\frac{1}{2\pi i} \oint_{\gamma_L} \! \frac{d \zeta_L}{\zeta_L} \int
 \! d^3 x d^4 \theta \mathcal{L}^{[2]}_L (z,
 \zeta_L)|_{\theta_{\perp} = 0} 
+ 
\frac{1}{2\pi i} \oint_{\gamma_R} \! \frac{d \zeta_R}{\zeta_R} \int
 \! d^3 x d^4 \theta \mathcal{L}^{[2]}_R (z,
 \zeta_R)|_{\theta_{\perp} = 0},
\end{eqnarray}
where the symbol $|_{\theta_{\perp} = 0}$ means that the 
superfields are projected on the $\mathcal{N} = 2$ superspace.

Classification of multiplets are similar to the $\mathcal{N} = 3$ case. 
Since the left and right parts have almost the same structure, 
let us focus on the left part for the moment.
A complex $\mathcal{O}(k)$ multiplet and a real $\mathcal{O}(-k,k)$ multiplet are
defined as (\ref{Ok_expansion}) and (\ref{Okk_expansion}), respectively.
Constraints for the components of a complex
$\mathcal{O}(k)$ multiplet $\Upsilon^{[n]}
=\displaystyle\sum_{l=0}^{k} \Upsilon_l \zeta^l$ 
are given by 
\begin{eqnarray}
\begin{aligned}
 & \bar{\mathbb{D}}_{\alpha} \Upsilon_0 = D_{\alpha}^{2 \bar{1}} \Upsilon_0 = 0, \\
 & \bar{\mathbb{D}}_{\alpha} \Upsilon_l = - D_{\alpha} ^{1 \bar{2}}
 \Upsilon_{l-1}, \quad D_{\alpha}^{2 \bar{1}} \Upsilon_l = \mathbb{D}_{\alpha}
 \Upsilon_{l-1}, \quad (1 \le l \le k), \\
 & \mathbb{D}_{\alpha} \Upsilon_k = D_{\alpha}^{1\bar{2}} \Upsilon_k = 0, 
\end{aligned}
\end{eqnarray}
while those on a real
$\mathcal{O} (-k,k)$ multiplet $U^{[2n]} = \displaystyle
\sum_{l=-k}^{k} U_l \zeta^l$ are 
\begin{eqnarray}
\begin{aligned}
 & \bar{\mathbb{D}}_{\alpha} U_{-k} = D_{\alpha}^{2 \bar{1}}
 U_{-k} = 0, \\
 & \bar{\mathbb{D}}_{\alpha} U_l = - D_{\alpha}^{1 \bar{2}}U_{l-1}, \quad
 D_{\alpha}^{2 \bar{1}} U_l = \mathbb{D}_{\alpha}
 U_{l-1},\quad (-k+1 \le l \le k), \\
 & D_{\alpha}^{1 \bar{2}} U_k = \mathbb{D}_{\alpha}U_k = 0.
\end{aligned} \label{G2-constN4}
\end{eqnarray}
The (ant)arctic multiplets and tropical multiplets are defined by 
taking $k\rightarrow \infty$ in the complex $\mathcal{O}(k)$ and the
real $\mathcal{O}(-k,k)$ multiplets, respectively.

We now consider the Chern-Simons 
action in $\mathcal{N} = 4$ projective superspace. 
Following the discussion in the previous section, we
consider the weight-2 Lagrangian in the left part, 
\begin{eqnarray}
\mathcal{L}^{(2)}_L = \frac{k}{8\pi} \mathcal{V}^{(0)}_{L} (z, v_L) G^{(2)}_{L} (z, v_L),
\end{eqnarray}
where $\mathcal{V}^{(0)}_{L}$ is the weight-0 tropical multiplet while $G^{(2)}_{L}$
is the gauge invariant weight-2 real $\mathcal{O}(-1,1)$ multiplet.
The constraints 
for the component superfields in $G^{(2)}_{L}$ 
are obtained from (\ref{G2-constN4}) as
\begin{eqnarray}
\begin{aligned}
 & \bar{\mathbb{D}}_{\alpha} \Phi_{L} = 0, \\
 & \mathbb{D}^2 L_{L} = \bar{\mathbb{D}}^2 L_{L} = 0, \\
 & \mathbb{D}_{\alpha} \bar{\Phi}_{L} = 0,
\end{aligned}
\end{eqnarray}
where they are projected on the $\mathcal{N}=2$ superspace.
Now we represent the gauge invariant multiplet $G_L^{(2)}$ by the gauge 
potential. The closed form of the solution to the $\mathcal{N} = 4$
projective superspace constraint was obtained in \cite{KuLiTa}.
The result is 
\begin{eqnarray}
G^{(2)}_L (v_L) &=& - \frac{i}{4} D^{(2) \bar{i} \bar{j}} \oint_{\gamma} \!
\frac{(v_R, d v_R)}{2\pi} \frac{u_{\bar{i}} u_{\bar{j}}}{(v_R, u_R)^2}
\mathcal{V}^{(0)}_R (v_R),
\end{eqnarray}
or equivalently
\begin{eqnarray}
G^{ij}_L = - \frac{i}{4} \oint_{\gamma} \frac{(v_R, dv_R)}{2\pi} 
D^{(-2) ij}
 \mathcal{V}^{(0)}_R (v_R).
\label{N4sol}
\end{eqnarray}
Here the left $\mathcal{O}(-1,1)$ multiplet is expanded as $G^{(2)}_L =
G^{ij}_L v_i v_j$.
We note that the left multiplet $G_L$ is represented by the right
multiplet $\mathcal{V}_R$ in the $\mathcal{N} = 4$ case.
Performing the contour integration, we find
\begin{equation}
G^{[2]}_L = 
\frac{i}{\zeta_L} 
\left(
- \frac{1}{4} \bar{\mathbb{D}}^{\alpha} \bar{\mathbb{D}}_{\alpha}
V_{R,1}
\right)
+ \frac{i}{2} \bar{\mathbb{D}}^{\alpha} \mathbb{D}_{\alpha} V_{R,0}
+ i \zeta_L 
\left(
\frac{1}{4} \mathbb{D}^{\alpha} \mathbb{D}_{\alpha} V_{R,-1}
\right),
\label{N4comp}
\end{equation}
where we have defined $G^{(2)}_L = (v^1)^2 (i \zeta_L) G^{[2]}_L$ and 
the component fields of $\mathcal{V}_R^{[0]}$
 in the Lindstr\"om-Ro\v{c}ek gauge:
\begin{eqnarray}
\mathcal{V}^{[0]}_R = \frac{1}{\zeta_R} V_{R,-1} + V_{R,0} + \zeta_R V_{R,1}.
\end{eqnarray}
The detailed calculations are found in Appendix B. 
The same analysis is applied to the right multiplet $G^{(2)}_R$. 
Therefore the $\mathcal{N} = 4$ Chern-Simons action in the $\mathcal{N} = 2$
superspace is given by 
\begin{eqnarray}
S_{CS} &=& 
\frac{ik}{16 \pi} 
\int d^3 x d^4\theta \ V_{L,0} \bar{\mathbb{D}}^{\alpha} \mathbb{D}_{\alpha} V_{R,0} 
+ \frac{ik}{8\pi} \int d^3 x d^2 \theta \ \Phi_L \Phi_R 
+ \frac{ik}{8\pi} \int d^3 x d^2 \bar{\theta} \ \bar{\Phi}_L \bar{\Phi}_R
\nonumber \\
& & + (L \leftrightarrow R).
\label{N4N2SS}
\end{eqnarray}
The calculation is the same with the $\mathcal{N} = 3$ case.

We note that there are two gauge potentials associated with the left and
right parts. From the construction, we find that 
each part is invariant under the two independent $U(1)$ gauge
transformations. Therefore the gauge group of the $\mathcal{N} = 4$
superconformal Chern-Simons action is interpreted as $U(1) \times
U(1)$. The gauge coupling constants for these gauge groups should be the same
due to the $SO(4)_R$ invariance. 
This situation is different from the $\mathcal{N} = 3$ case where
we can construct the gauge invariant action with a single $U(1)$
gauge group.
We need more than one gauge potential to construct $\mathcal{N} \ge
4$ supersymmetric Chern-Simons theories.
There is a natural interpretation for this gauge group enhancement. 
It has been discussed that the $\mathcal{N} \ge 4$ supersymmetry
completion of a pure Chern-Simons term is impossible provided that
there is only one gauge potential. One needs multiple gauge
potentials to make the action be invariant under $\mathcal{N} \ge 4$
supersymmetries \cite{NiGa, NiGa2, BrGa}. We therefore have to introduce more than
single gauge group or may need 3-algebra structure which enable the gauge
potential to have multiple components (with 3-algebra indices)
\cite{Ra, HoLeLeLePa, AhBeJaMa, GaWi}. 

For the matter part, we can construct the $\mathcal{N} = 4$ action which 
includes $U(1)$ gauge interactions.
The matter multiplets are introduced as 
weight-1 left (and right) (ant)arctic
multiplets $\Upsilon^{[1]}$.
Following the Chern-Simons case, since the left and right parts are almost the same,
we focus on the left part. We define the gauge covariant derivative as 
\begin{eqnarray}
\mathcal{D}_{\alpha}^{[1] \bar{k}} 
= e^{-q \mathcal{V}_{-}} D_{\alpha}^{[1]\bar{k}} e^{q \mathcal{V}_{-}} 
= e^{q \mathcal{V}_{+}} D_{\alpha}^{[1] \bar{k}} e^{-q \mathcal{V}_{+}},
\end{eqnarray}
where we have decomposed the left $U(1)$ vector multiplet 
$\mathcal{V}^{[0]}_{L}=\mathcal{V}_++\mathcal{V}_{-}$ in the same
way as the $\mathcal{N} = 3$ case.
We can show that the matter multiplets $\tilde{\Upsilon}^{[1]} = e^{\mathcal{V}_{+}}
\Upsilon^{[1]}$ and $\tilde{\bar{\Upsilon}}^{[1]} = e^{\mathcal{V}_{-}}
\bar{\Upsilon}^{[1]}$ satisfy the gauge covariantized projective
constraints:
\begin{eqnarray}
\mathcal{D}^{[1] \bar{k}}_{\alpha} \tilde{\Upsilon}^{[1]} = \mathcal{D}_{\alpha}^{[1]
 \bar{k}} \tilde{\bar{\Upsilon}}^{[1]} = 0.
\label{N4-matter_constraints}
\end{eqnarray}
Using the anticommutation relations of the gauge covariant derivatives
in Appendix C and the constraints \eqref{N4-matter_constraints}, one can
construct the matter action together with the right part 
as in the same procedure in $\mathcal{N} = 3$ case. 

Finally we comment on the classification of the multiplets and 
superconformal invariant action in the $\mathcal{N} = 4$ projective
superspace. In this paper, we consider the left and right projective
multiplets. 
They can be defined independently and the superconformal
invariant action \eqref{N4action} is just the sum of the left and right
parts. However, one can consider multiplets constructed by the products
of the left and right multiplets, called hybrid projective multiplets
with weight $(n,m)$:
\begin{equation}
Q^{(n,m)} = Q^{(n)}_L (v_L) Q^{(m)}_R (v_R),
\end{equation}
where $Q^{(n)}_L$ ($Q^{(m)}_R$) is a left (right) projective multiplet
with weight $n$ ($m$).
The hybrid projective multiplets satisfy the analyticity condition determined by the new covariant
derivative $D^{(1,1)}_{\alpha} = v_i v_{\bar{k}} D^{i
\bar{k}}_{\alpha}$:
\begin{equation}
D^{(1,1)}_{\alpha} Q^{(n,m)} = 0.
\end{equation}
The $\mathcal{N} = 4$ superconformal 
invariant action constructed by the hybrid projective multiplets is
discussed in \cite{KuLiTa}.
It would be interesting to study $\mathcal{N} \ge 4$ Chern-Simons-matter
models by using the hybrid projective multiplets. We leave this
possibility to future works.

\section{Conclusion and discussions}
We have constructed the $\mathcal{N} = 3$ superconformal Chern-Simons matter theories
with Abelian gauge group in the three-dimensional projective
superspace. The weight-2 Lagrangian is given by the
product of the weight-0 tropical multiplet $\mathcal{V}^{(0)}$ and the
gauge invariant weight-2 $\mathcal{O}(-1,1)$ multiplet $G^{(2)}$. 
We have solved the constraints for the $\mathcal{O}(-1,1)$ multiplet and
expressed its component superfields by 
those of the tropical multiplet.
We have also constructed the matter action interacting with the gauge
fields. The gauge covariant derivatives are defined by using the tropical
multiplet and the matter (ant)arctic multiplets satisfy the gauge
covariantized projective constraints.

These constructions of the actions are generalized to the $\mathcal{N} = 4$ cases. 
In order to introduce the R-symmetry group $SO(4)_R \sim SU(2) \times
SU(2)/\mathbb{Z}_2$, we have introduced the mirror $\mathbb{C}P^1$s and
considered the $\mathcal{N} = 4$ superconformal projective superspace.
The Chern-Simons and the matter actions are constructed as in the similar
way for the $\mathcal{N} = 3$ cases. 
We need to introduce two vector potentials for the $\mathcal{N} = 4$
superconformal invariant action. This fact leads to the result that the 
gauge group of the theory is actually the product of the two groups, namely,
$U(1) \times U(1)$. 

A few comments on the non-Abelian generalizations of our constructions 
are in order. 
Let us try to find the $\mathcal{N} = 3$ superconformal Chern-Simons-matter
action with non-Abelian gauge groups in the projective superspace. 
A natural way for the non-Abelian generalization of the $\mathcal{N} = 3$ 
Chern-Simons action \eqref{CSLagrangian} is to replace the projective
multiplets $\mathcal{V}^{(0)}$ and $G^{(2)}$ by the ones with adjoint
representations of a non-Abelian group. 
The Lagrangian may be given by 
\begin{eqnarray}
 \mathcal{L}^{(2)}= \frac{k}{8\pi}{\rm Tr}
\left[
\mathcal{V}^{(0)}(z,v)G^{(2)}(z,v)
\right], 
\label{NACS}
\end{eqnarray}
where the trace is taken over the adjoint representation of the gauge group.
As in the Abelian case, 
one may try to find solutions to the projective superspace constraints
$D^{(2)}_{\alpha} G^{(2)} = 0$ and express the 
$\mathcal{N}=2$ components 
in $G^{(2)}$ in terms of those in $\mathcal{V}^{(0)}$.
Using this solution, the Lagrangian \eqref{NACS} 
should reproduce the $\mathcal{N} = 3$ Chern-Simons action
(\ref{N3CS}) in $\mathcal{N} = 2$ superspace.
However, we find that the straightforward generalization 
of the Abelian solution \eqref{o2-comp}
 to the non-Abelian case does not work. 
Even more, the complicated dependence on the auxiliary variable $t$ in (\ref{N3CS}) 
should be incorporated with the action in the projective superspaces.

The formulation of non-Abelian Chern-Simons theories in the 
$\mathcal{N} = 4$ projective superspace is also an interesting problem.
The $\mathcal{N} \ge 4$ superconformal Chern-Simons theories with
bi-fundamental matters have been studied intensively 
in the context of the world-volume effective theories of M2-branes.
In order to construct the $\mathcal{N} = 4$ superconformal action for the matter
fields with the bi-fundamental representation, one may need to introduce
the hybrid projective multiplets. This is because the bi-fundamental
matters should couple two gauge potentials simultaneously. 
We explore these possibilities in the future works.

Finally, let us comment on the construction 
of the $\mathcal{N}=4$ Chern-Simons-matter model
discussed in \cite{ChDoSa}.
In \cite{ChDoSa}, manifestly $\mathcal{N}=4$ supersymmetric 
construction of the Bagger-Lambert-Gustavsson type action 
was studied. 
The action is based on the 3-algebra gauge invariance 
and is formulated in the $\mathcal{N} = 4$ projective superspace.
This $\mathcal{N} = 4$ projective superspace is defined by the
dimensional reduction of the one in four dimensions and
$SO(4)_R \sim SU(2) \times SU(2)/\mathbb{Z}_2$ R-symmetry is not manifest.
There are only one matter multiplet and one vector multiplet 
in the action and the mirror pairs (left or right multiplets) do not
exist, which are necessary to keep $\mathcal{N}=4$ superconformal invariance. 
Consequently, it is unclear whether their action is $\mathcal{N}=4$ 
superconformal invariant or not. 
It would be interesting to formulate $\mathcal{N} \ge 4$ 
Chern-Simons-matter theories based
on the 3-algebra gauge groups. Our construction may be applicable to the
3-algebra gauge groups. We hope we come back to this problem in the
future researches.

\subsection*{Acknowledgments}
The hospitality of the University of Western Australia during the first stage of this 
project is gratefully acknowledged. M.A. thanks S.M. Kuzenko for discussion and giving
useful comments.
The authors would like to thank Y.~Imamura for giving us a comment on
pure supersymmetric Chern-Simons theories. 
The work of M.~A. is supported in part by the Research Program 
MSM6840770029 and by the project of International Cooperation ATLAS-CERN 
of the Ministry of Education, Youth and Sports of the Czech Republic.
The work of S.~S. is supported by the Research Fellowship of the Japan Society 
for the Promotion of Science (JSPS).

\begin{appendix}
\section{
Conventions and notations in three-dimensional superspaces
}\label{appendixA}
This appendix provides the basic conventions 
and notations of ordinary superspaces
in three dimensions.
We use the convention of the three-dimensional metric $\eta_{mn} = 
\mathrm{diag} (-1,+1,+1)$ with $m, n$ run from $0$ to $2$.
The three-dimensional $\mathcal{N} = 2$ superspace is represented by the 
coordinate $z^A=(x^{m}, \theta^{\alpha}, \bar{\theta}^{\alpha})$ where 
$\theta, \bar{\theta}$ are two component spinors. The gamma matrices are defined by 
$(\gamma^{m})_{\alpha} {}^{\beta} = (i \tau^2, \tau^1 ,\tau^3)$ 
which satisfies $\{\gamma^{m}, \gamma^{n} \} = 2 \eta^{mn}$.
$\tau^I \ (I=1,2,3)$ are the Pauli matrices.
The spinor indices are raised and lowered by the anti-symmetric 
symbol $\varepsilon^{12} = - \varepsilon_{12} = 1$. 
The supercovariant derivatives in $\mathcal{N} = 2$ superspace 
are defined by 
\begin{eqnarray}
\mathbb{D}_{\alpha} = \partial_{\alpha} + i (\gamma^{m} \bar{\theta})_{\alpha} 
\partial_{m}, \qquad 
\bar{\mathbb{D}}_{\alpha} = - \bar{\partial}_{\alpha} - i (\theta 
\gamma^{m})_{\alpha} \partial_{m}. \label{cov1}
\end{eqnarray}
These satisfy the following relations
\begin{eqnarray}
\{\mathbb{D}_{\alpha}, \bar{\mathbb{D}}_{\beta} \} = - 2 i
 \gamma^{m}_{\alpha \beta} \partial_{m}, \quad \{ \mathbb{D}_{\alpha},
 \mathbb{D}_{\beta} \} = \{\bar{\mathbb{D}}_{\alpha},
 \bar{\mathbb{D}}_{\beta} \} = 0.
\end{eqnarray}
The Grassmann measure of integration is defined by 
\begin{eqnarray}
d^2 \theta = - \frac{1}{4} d \theta^{\alpha} d \theta_{\alpha}, \quad 
d^2 \bar{\theta} = - \frac{1}{4} d \bar{\theta}^{\alpha} d 
\bar{\theta}_{\alpha}, \quad d^4 \theta = d^2 \theta d^2 \bar{\theta},
\end{eqnarray}
such that they satisfy
\begin{eqnarray}
\int \! d^2 \theta \ \theta^2 = 1, \quad \int \! d^2 \bar{\theta} \ 
\bar{\theta}^2=1, \quad \int \! d^4 \theta \ \theta^2 \bar{\theta}^2 = 1.
\end{eqnarray}
Within the space-time integration, the following relation holds,
\begin{eqnarray}
\int \! d^4 \theta \ F(z) = \left. \frac{1}{16} (\mathbb{D}^2 \bar{\mathbb{D}}^2 F(z)) 
\right|_{\theta = \bar{\theta} = 0},
\end{eqnarray}
where $F(z)$ is a superfield. 
The chiral and anti-chiral coordinates are defined by 
\begin{eqnarray}
x^{m}_L = x^{m} + i \theta \gamma^{m} \bar{\theta}, \qquad 
x^{m}_R = x^{m} - i \theta \gamma^{m} \bar{\theta}.
\end{eqnarray}
\end{appendix}

We use the following relations among $\mathcal{N} = 2$, 
$\mathcal{N} = 3$ and $\mathcal{N} = 4$ superspaces \cite{KuPaTaUn}:
\begin{eqnarray}
& & \theta^{\alpha} = \theta^{\alpha}_{11} = \theta^{\alpha}_{1\bar{1}},
 \quad 
\bar{\theta}^{\alpha} = \theta^{\alpha}_{22} =
\theta^{\alpha}_{2\bar{2}}, \\
& & \mathbb{D}_{\alpha} = D^{11}_{\alpha} = D^{1 \bar{1}}_{\alpha},
 \quad 
\bar{\mathbb{D}}_{\alpha} = - D^{22}_{\alpha} = - D^{2\bar{2}}_{\alpha}.
\end{eqnarray}
The $\mathcal{N} = 2$ projection of $\mathcal{N} = 3, 4$ superfields
$\Phi (z^A)$ is defined by 
\begin{eqnarray}
\left. \Phi \right| = \left. \Phi (z^A) \right|_{\theta_{\bot} = 0},
\end{eqnarray}
where $\theta_{\bot}$ is $\theta_{12}$ ($\mathcal{N} = 3$) or
$\theta_{1\bar{2}}, \theta_{2\bar{1}}$ ($\mathcal{N} = 4$).

\section{Solutions to the constraint}
In this section, the detailed calculations of the integral in
\eqref{N3sol} and its $\mathcal{N} = 4$ counterpart \eqref{N4sol} are
shown. 

We fix $u_i = (1,0)$. The integral measure is $v^i d v_i =
- (v^1)^2 d \zeta$ and the products of the supercovariant derivatives are
\begin{eqnarray}
\begin{aligned}
& D^{(-2)\alpha} D^{(-2)}_{\alpha} = \frac{1}{(v^1)^2}
(- \zeta \mathbb{D}^2 + \mathbb{D} D^{12}), \\
& D^{(-2) \alpha} D^{(0)}_{\alpha} = \frac{1}{(v^1)^2}
\left(
- \frac{1}{2} \zeta \mathbb{D}^2 - \frac{1}{2\zeta} \mathbb{D} \bar{\mathbb{D}}
\right), \\
& D^{(0)\alpha} D^{(0)}_{\alpha} = \zeta^2 \mathbb{D}^2 - 2 \zeta
 \mathbb{D} D^{12} + (D^{12})^2.
\end{aligned}
\label{SD2}
\end{eqnarray}
One can eliminate $D^{12}$ in each expression using the relation 
$D^{12} = - \frac{1}{2\zeta} (D^{[2]} + \bar{\mathbb{D}} - \zeta^2 \mathbb{D})$.
We start from the first term in \eqref{N3sol}:
\begin{eqnarray}
& & - \frac{1}{8 \pi i} \oint_{\gamma} (v, dv) 
\left(
- \frac{i}{2} 
\right)
(w, v)^2 (D^{(-2)})^2 \mathcal{V}^{(0)} (v) 
\nonumber \\
&=& - \frac{i}{2} \frac{1}{8 \pi i} 
\oint_{\gamma} d \zeta_v \ 
(w_1^2 + 2 \zeta_v w_1 w_2 + \zeta_v^2 w_2^2)
\mathbb{D}^2 
\left(
\frac{1}{\zeta_v} V_{-1} + V_0 + \zeta_v V_1 
\right) 
\nonumber \\
&=& - \frac{i}{8} w_1^2 \mathbb{D}^2 V_{-1},
\end{eqnarray}
where we have used the projective superspace constraint 
$D^{[2]} \mathcal{V}^{[0]} = 0$ and defined $\zeta_v = \frac{v^2}{v^1}$.
The second term in \eqref{N3sol} is 
\begin{eqnarray}
& & - \frac{1}{8 \pi i} \oint_{\gamma} (v, dv) \ 
8 \frac{(w,v)(w,u)}{(v,u)} D^{(-2)} D^{(0)}
\mathcal{V}^{(0)} 
\nonumber \\
&=& \frac{1}{\pi i} \oint_{\gamma} \! d \zeta_v \ (w_1 w_2 + \zeta_v
 w_2^2) 
\left(
- \frac{1}{2} \zeta_v \mathbb{D}^2 - \frac{1}{2 \zeta_v} \mathbb{D} \bar{\mathbb{D}}
\right)
\left(
\frac{1}{\zeta_v} V_{-1} + V_0 + \zeta_v V_1
\right) 
\nonumber \\
&=& w_1 w_2 \mathbb{D} \bar{\mathbb{D}} V_0.
\end{eqnarray}
The third term is 
\begin{eqnarray}
& & - \frac{1}{8 \pi i} \oint_{\gamma} (v,dv) \ 
2i \frac{(w,u)^2}{(v,u)^2} (D^{(0)})^2 \mathcal{V}^{(0)}
\nonumber \\
&=& \frac{1}{4\pi} \oint_{\gamma} \! d \zeta_v \
(w_2)^2 \frac{1}{4 \zeta_v^2} \bar{\mathbb{D}}^2 
\left(
\frac{1}{\zeta_v} V_{-1} + V_0 + \zeta_v V_1
\right)
\nonumber \\
&=& \frac{i}{2} (w_2)^2 \bar{\mathbb{D}}^2 V_1.
\end{eqnarray}
From these results we find
\begin{eqnarray}
G^{[2]} (w) = \frac{i}{\zeta_w} 
\left(
\frac{i}{8} \bar{\mathbb{D}}^2 V_1 
\right)
+ i \bar{\mathbb{D}} \mathbb{D} V_0
+ i \zeta_w 
\left(
- \frac{i}{8} \mathbb{D}^2 V_{-1}
\right).
\end{eqnarray}
Therefore \eqref{o2-comp} is obtained.

Next, we calculate the integral \eqref{N4sol}. 
The left $\mathcal{O}(-1,1)$ multiplet $G^{(2)}_L = G^{ij} v_i v_j$ is expanded as 
\begin{eqnarray}
G^{(2)}_L &=& (v^1)^2 (i \zeta_L) 
\left(
- i \zeta_L G^{11} + 2 i G^{12} - i \zeta^{-1}_L G^{22}
\right) \nonumber \\
&\equiv& (v^1)^2 (i\zeta_L) G^{[2]}_L.
\end{eqnarray}
Then, we find
\begin{eqnarray}
G^{11} &=& \frac{1}{4} \oint_{\gamma} \frac{(v^{\bar{1}})^2 d \zeta_R}{2
 \pi i} \frac{1}{(v^{\bar{1}})^2} (D^{1\bar{1}})^2 
\left(
\frac{1}{\zeta_R} V^{R}_{-1} + V^R_0 + \zeta_R V^R_1
\right) 
\nonumber \\
&=& \frac{1}{4} \mathbb{D}^2 V^R_{-1}.
\end{eqnarray}
Similarly, the other components are 
\begin{eqnarray}
G^{12} &=& \frac{1}{4} \oint_{\gamma} \frac{d \zeta_R}{2\pi i} 
D^{1\bar{1}} D^{1\bar{2}} 
\left(
\frac{1}{\zeta_R} V^{R}_{-1} + V^R_0 + \zeta_R V^R_1
\right)
\nonumber \\
&=& - \frac{1}{4} \bar{\mathbb{D}} \mathbb{D} V^R_0, \\
G^{22} &=& \frac{1}{4} \oint_{\gamma} \frac{d \zeta_R}{2\pi i}
 (D^{1\bar{2}})^2 
\left(
\frac{1}{\zeta_R} V^{R}_{-1} + V^R_0 + \zeta_R V^R_1
\right)
\nonumber \\
&=& \frac{1}{4} \mathbb{D}^2 V^R_1,
\end{eqnarray}
where we have used the $\mathcal{N} = 4$ projective superspace
constraints on $\mathcal{V}_R^{(0)}$. 
Finally we obtain \eqref{N4comp}.

\section{Anticommutation relations of supercovariant derivatives}
The $\mathcal{N} = 3$ super gauge covariant derivatives satisfy the following
anticommutation relations
\begin{eqnarray}
\begin{aligned}
\{ \mathcal{D}_{\alpha}, \mathcal{D}_{\beta} \} =& 
0, \\
\{\bar{\mathcal{D}}_{\alpha}, \bar{\mathcal{D}}_{\beta} \} 
=& 2 \bar{\mathbb{D}}_{(\alpha} D^{12}_{\beta)} V_{-1}, \\
\{ \mathcal{D}_{\alpha}^{12}, \mathcal{D}_{\beta}^{12} \} =& 
\{D^{12}_{\alpha}, D^{12}_{\beta} \} + \frac{1}{2} \{D^{12}_{\alpha},
D^{12}_{\beta} \} V_0 - \frac{1}{2} D^{12}_{(\alpha} \mathbb{D}_{\beta)}
V_{-1}, \\
\{\mathcal{D}_{\alpha}, \mathcal{D}_{\beta}^{12} \} =& - \frac{1}{2}
 \mathbb{D}_{\alpha} \mathbb{D}_{\beta} V_{-1}, \\
\{\mathcal{D}_{\alpha}, \bar{\mathcal{D}}_{\beta} \} =& 
\{\mathbb{D}_{\alpha}, \bar{\mathbb{D}}_{\beta} \} + 
\frac{1}{2} \{\mathbb{D}_{\alpha}, \bar{\mathbb{D}}_{\beta} \} V_0 + 2
\mathbb{D}_{\alpha} D^{12}_{\beta} V_{-1}, \\
\{\bar{\mathcal{D}}_{\alpha}, \mathcal{D}_{\beta}^{12}\} =& -
 \frac{1}{2} \bar{\mathbb{D}}_{\alpha} \mathbb{D}_{\beta} V_{-1} + 2
 D^{12}_{\beta} D^{12}_{\alpha} V_{-1},
\end{aligned}
\label{B1}
\end{eqnarray}
where we have taken $q = 1$ for simplicity. 

For the $\mathcal{N} = 4$ case, the gauge covariant derivative 
in the left sector is expanded as 
\begin{eqnarray}
\mathcal{D}^{[1]\bar{k}}_{\alpha} = \mathcal{D}^{2 \bar{k}}_{\alpha} -
 \zeta \mathcal{D}^{1 \bar{k}}_{\alpha},
\end{eqnarray}
where we have defined
\begin{eqnarray}
\mathcal{D}^{2 \bar{k}}_{\alpha} &=& D^{2 \bar{k}}_{\alpha} + \frac{q}{2}
 D^{2 \bar{k}}_{\alpha} V_0 - q D^{1 \bar{k}}_{\alpha} V_{-1}, \\
\mathcal{D}^{1 \bar{k}}_{\alpha} &=& D^{1 \bar{k}}_{\alpha} +
 \frac{q}{2} D^{1 \bar{k}}_{\alpha} V_0.
\end{eqnarray}
These satisfy the following anticommutation relations with $q=1$:
\begin{eqnarray}
\begin{aligned}
 & \{ \mathcal{D}_{\alpha}^{1 \bar{1}}, \mathcal{D}_{\beta}^{1 \bar{1}} \}
 = 0, \\
 & \{ \mathcal{D}_{\alpha}^{1 \bar{1}}, \mathcal{D}_{\beta}^{1 \bar{2}} \}
 = 0, \\
 & \{ \mathcal{D}_{\alpha}^{1 \bar{1}}, \mathcal{D}_{\beta}^{2 \bar{1}} \}
 = - \mathbb{D}_\alpha\mathbb{D}_\beta V_{-1}, \\
 & \{ \mathcal{D}_{\alpha}^{1 \bar{1}}, \mathcal{D}_{\beta}^{2 \bar{2}} \}
 = - \{\mathbb{D}_{\alpha}, \bar{\mathbb{D}}_{\beta} \} - \frac{1}{2}
 \{\mathbb{D}_{\alpha}, \bar{\mathbb{D}}_{\beta} \} V_0 -
 \mathbb{D}_{\alpha} D^{1 \bar{2}}_{\beta} V_{-1}, \\
 & \{ \mathcal{D}_{\alpha}^{1 \bar{2}}, \mathcal{D}_{\beta}^{1 \bar{2}} \}
 = 0
, \\
 & \{ \mathcal{D}_{\alpha}^{1 \bar{2}}, \mathcal{D}_{\beta}^{2 \bar{1}} \}
 = \{D^{1 \bar{2}}_{\alpha}, D^{2 \bar{1}}_{\beta} \} + 
\frac{1}{2} \{D^{1 \bar{2}}_{\alpha}, D^{2 \bar{1}}_{\beta} \} V_0 -
D^{1 \bar{2}}_{\alpha} \mathbb{D}_{\beta} V_{-1}, \\
 & \{ \mathcal{D}_{\alpha}^{1 \bar{2}}, \mathcal{D}_{\beta}^{2 \bar{2}} \}
 = -D^{1\bar{2}}_{\alpha} D^{1\bar{2}}_{\beta}V_{-1}, \\
 & \{ \mathcal{D}_{\alpha}^{2 \bar{1}}, \mathcal{D}_{\beta}^{2 \bar{1}} \}
 = -D^{2\bar{1}}_{(\alpha}\mathbb{D}_{\beta)}V_{-1}
, \\
 & \{ \mathcal{D}_{\alpha}^{2 \bar{1}}, \mathcal{D}_{\beta}^{2 \bar{2}} \}
 = -D^{2\bar{1}}_{\alpha} D^{1\bar{2}}_{\beta}V_{-1}+\bar{\mathbb{D}}_\beta\mathbb{D}_\alpha V_{-1}, \\
 & \{ \mathcal{D}_{\alpha}^{2 \bar{2}}, \mathcal{D}_{\beta}^{2 \bar{2}} \}
 = \bar{\mathbb{D}}_{(\alpha}D^{1\bar{2}}_{\beta)}V_{-1}.
\end{aligned}
\label{B2}
\end{eqnarray}
Similar definition of the gauge covariant derivative is applied to the
right sector.


\begin{thebibliography}{0}
%
\bibitem{KaLe}
  H.~-C.~Kao, K.~-M.~Lee,
  Phys.\ Rev.\  {\bf D46 } (1992)  4691
  [hep-th/9205115].
%
\bibitem{NiGa}
  H.~Nishino, S.~J.~Gates, Jr.,
  Int.\ J.\ Mod.\ Phys.\  {\bf A8 } (1993)  3371.
%
\bibitem{NiGa2}
  H.~Nishino and S.~J.~J.~Gates,
  Nucl.\ Phys.\  B {\bf 480} (1996) 573
  [arXiv:hep-th/9606090].
%
\bibitem{BrGa}
  R.~Brooks and S.~J.~J.~Gates,
  Nucl.\ Phys.\  B {\bf 432} (1994) 205
  [arXiv:hep-th/9407147].
%
\bibitem{BaLa}
  J.~Bagger and N.~Lambert,
  Phys.\ Rev.\  D {\bf 75} (2007) 045020
  [hep-th/0611108], \\
  Phys.\ Rev.\  D {\bf 77} (2008) 065008
  [arXiv:0711.0955 [hep-th]], \\
  JHEP {\bf 0802} (2008) 105
  [arXiv:0712.3738 [hep-th]].
%
\bibitem{Gu}
  A.~Gustavsson,
  Nucl.\ Phys.\ B {\bf 811} (2009) 66
  [arXiv:0709.1260 [hep-th]].
%
\bibitem{Ra}
  M.~Van Raamsdonk,
  JHEP {\bf 0805} (2008) 105
  [arXiv:0803.3803 [hep-th]].
%
\bibitem{HoLeLeLePa}
  K.~Hosomichi, K.~-M.~Lee, S.~Lee, S.~Lee, J.~Park,
  JHEP {\bf 0807 } (2008)  091 
  [arXiv:0805.3662 [hep-th]], \\
  JHEP {\bf 0809 } (2008)  002 
  [arXiv:0806.4977 [hep-th]].
%
\bibitem{AhBeJaMa}
  O.~Aharony, O.~Bergman, D.~L.~Jafferis and J.~Maldacena,
  JHEP\ {\bf 0810} (2008) 091
  [arXiv:0806.1218 [hep-th]]. 
%
\bibitem{Iv}
  E.~A.~Ivanov,
  Phys.\ Lett.\  B {\bf 268} (1991) 203.
%
\bibitem{GaNi1}
  S.~J.~J.~Gates and H.~Nishino,
  Phys.\ Lett.\  B {\bf 281} (1992) 72.
%
\bibitem{Ce}
  M.~Cederwall,
  JHEP {\bf 0809 } (2008)  116 
  [arXiv:0808.3242 [hep-th]], \\
  JHEP {\bf 0810 } (2008)  070 
  [arXiv:0809.0318 [hep-th]].
%
\bibitem{GaIvOgSo}
  A.~Galperin, E.~Ivanov, V.~Ogievetsky, E.~Sokatchev,
  JETP Lett.\  {\bf 40 } (1984)  912-916.
%
\bibitem{GaIvKaOgSo}
  A.~Galperin, E.~Ivanov, S.~Kalitsyn, V.~Ogievetsky, E.~Sokatchev,
  Class.\ Quant.\ Grav.\  {\bf 1 } (1984)  469-498.
%
\bibitem{HoLe}
  P.~S.~Howe, M.~I.~Leeming,
  Class.\ Quant.\ Grav.\  {\bf 11 } (1994)  2843-2852 
  [hep-th/9408062].
%
\bibitem{Zu}
  B.~M.~Zupnik,
  Phys.\ Lett.\  {\bf B660 } (2008)  254-259.
  [arXiv:0711.4680 [hep-th]], \\
  Theor. \ Math. \ Phys. {\bf 157} (2008) 1550
  [arXiv:0802.0801 [hep-th]].
%
\bibitem{BuIvLePlSaZu}
  I.~L.~Buchbinder, E.~A.~Ivanov, O.~Lechtenfeld, N.~G.~Pletnev, I.~B.~Samsonov, B.~M.~Zupnik,
  JHEP {\bf 0903 } (2009)  096
  [arXiv:0811.4774 [hep-th]].
%
\bibitem{KaLiRo}
  A.~Karlhede, Lindstr\"om and M.~Ro\v{c}ek,
  Phys.\ Lett.\ {\bf B147 } (1984) 297.
%
\bibitem{LR1}
  U.~Lindstr\"om, M.~Ro\v{c}ek,
  Commun.\ Math.\ Phys.\  {\bf 115} (1988) 21.
%
\bibitem{LiRo}
  U.~Lindstr\"om, M.~Ro\v{c}ek,
  Commun.\ Math.\ Phys.\  {\bf 128 } (1990)  191.
%
\bibitem{Ku}
  S.~M.~Kuzenko,
  Int.\ J.\ Mod.\ Phys.\  A {\bf 14} (1999) 1737
  [hep-th/9806147].
%
\bibitem{KuPaTaUn}
S.~M.~Kuzenko, J.~H.~Park, G.~Tartaglino-Mazzucchelli and R.~Unge,
JHEP {\bf 1101} (2011) 146
[arXiv:1011.5727 [hep-th]].
%
\bibitem{KuLiTa}
S.~M.~Kuzenko, U.~Lindstr\"om and G.~Tartaglino-Mazzucchelli,
JHEP {\bf 1103} (2011) 120
[arXiv:1101.4013 [hep-th]].
%
\bibitem{ChDoSa}
  S.~Cherkis, V.~Dotsenko and C.~Saemann,
  Phys.\ Rev.\  D {\bf 79} (2009) 086002
  [arXiv:0812.3127 [hep-th]].
%
\bibitem{Sc}
  J.~H.~Schwarz,
  JHEP {\bf 0411} (2004) 078
  [hep-th/0411077].
%
\bibitem{GaYi}
  D.~Gaiotto, X.~Yin,
  JHEP {\bf 0708 } (2007)  056 
  [arXiv:0704.3740 [hep-th]].
%
\bibitem{GaWi}
  D.~Gaiotto and E.~Witten,
  JHEP {\bf 1006} (2010) 097
  [arXiv:0804.2907 [hep-th]].
%
%
\bibitem{KuLi}
  S.~M.~Kuzenko and W.~D.~I.~Linch,
  JHEP {\bf 0602} (2006) 038
  [hep-th/0507176].
%
\bibitem{Ku2}
  S.~M.~Kuzenko,
  Phys.\ Lett.\  B {\bf 644} (2007) 88
  [hep-th/0609078]. 
%
\bibitem{GoUn}
  F.~Gonzalez-Rey and R.~von Unge,
  Nucl.\ Phys.\ B\ {\bf 516} (1998) 449
  [hep-th/9711135].
\end{thebibliography}
\end{document}